# Cloudy-Maraston: integrating nebular continuum and line emission with the Maraston stellar population synthesis models

Sophie L. Newman ,[1★] Christopher C. Lovell ,[1] Claudia Maraston,[1] William J. Roper ,[2] Aswin P. Vijayan,[2] Stephen M. Wilkins ,[2] Mauro Giavalisco[3] and Aayush Saxena [4]

[1]*Institute of Cosmology & Gravitation, University of Portsmouth, Portsmouth, PO1 3FX, UK*
[2]*Astronomy Centre, University of Sussex, Falmer, Brighton, BN1 9QH, UK*
[3]*Department of Astronomy, University of Massachusetts, Amherst, MA 01003, USA*
[4]*Department of Physics, University of Oxford, Oxford, OX1 3RH, UK*

Accepted 2025 October 27. Received 2025 October 27; in original form 2025 January 6

## ABSTRACT

The *James Webb Space Telescope* has ushered in an era of abundant high-redshift observations of young stellar populations characterized by strong emission lines, motivating us to integrate nebular emission into the new Maraston stellar population model which incorporates the latest Geneva stellar evolutionary tracks for massive stars with rotation. We use the photoionization code CLOUDY to obtain the emergent nebular continuum and line emission for a range of modelling parameters, then compare our results to observations on various emission line diagnostic diagrams. We carry out a detailed comparison with several other models in the literature assuming different input physics, including modified prescriptions for stellar evolution and the inclusion of binary stars, and find close agreement in the H $\beta$, H $\alpha$, [N II]$\lambda$6583, and [S II]$\lambda$6716, 6731 luminosities between the models. However, we find significant differences in lines with high ionization energies, such as He II$\lambda$1640 and [O III]$\lambda$5007, due to large variations in the hard ionizing photon production rates. The models differ by a maximum of $\Delta\hat{Q}_{[\mathrm{OIII}]\lambda5007} \approx 10^{44}\ \mathrm{s^{-1}\ M_{\odot}^{-1}}$, where these differences are mostly caused by the assumed stellar rotation and effective temperatures for the Wolf Rayet phase. Interestingly, rotation and uncorrected effective temperatures in our single star population models alone generate [O III] ionizing photon production rates higher than models including binary stars with ages between 1 to 6 Myr. These differences highlight the dependence of derived properties from SED fitting on the assumed model, as well as the sensitivity of predictions from cosmological simulations.

## 1 INTRODUCTION

Stellar population synthesis (SPS) models, traditionally called evolutionary population synthesis models, are fundamental tools in deciphering the complex story of galaxy evolution (e.g. B. M. H. Tinsley 1967; L. Searle, W. L. W. Sargent & W. G. Bagnuolo 1973; G. Bruzual A. 1983; G. Worthey 1994; M. Fioc & B. Rocca-Volmerange 1997; C. Leitherer et al. 1999; G. Bruzual & S. Charlot 2003; D. Thomas, C. Maraston & R. Bender 2003; C. Maraston 2005; C. Conroy, J. E. Gunn & M. White 2009a; J. J. Eldridge et al. 2017; I. Millán-Irigoyen et al. 2021). By simulating the aggregate properties of coeval and composite populations these models offer insights into the ages (e.g. D. Thomas et al. 2005; C. Maraston et al. 2006; P. Serra & S. C. Trager 2007; M. Koleva et al. 2008; I. Ferreras et al. 2009; J. Miner, J. A. Rose & G. Cecil 2011; R. Tojeiro et al. 2011), initial mass function (IMF, e.g. I. K. Baldry & K. Glazebrook 2003; C. Conroy & P. G. van Dokkum 2012), masses (e.g. S. L. Finkelstein et al. 2010; E. Curtis-Lake et al. 2013), and formation history (e.g. C. Papovich, M. Dickinson & H. C. Ferguson 2001; S. Salim et al. 2007; C. Maraston et al. 2010; S. Wuyts et al. 2011) of stellar populations. Populations of young stars typically generate significant line emission due to the hard ionizing radiation emitted by their constituent massive stars, which ionizes the surrounding birth cloud. There has been an abundance of spectroscopic observations of these young stellar populations at high redshift by the *James Webb Space Telescope* (*JWST*; e.g. S. L. Finkelstein et al. 2022; C. M. Casey et al. 2023; D. J. Eisenstein et al. 2023) showing strong emission lines (e.g. A. J. Bunker et al. 2023; A. J. Cameron et al. 2023; S. Tacchella et al. 2023; K. Boyett et al. 2024; M. Curti et al. 2024); to interpret these results, the self-consistent treatment of emission lines with SPS models is needed.

To interpret the photometric and spectroscopic observations from the epoch of reionization captured by *JWST* (e.g. D. Schaerer et al. 2022; A. C. Carnall et al. 2023; J. A. A. Trussler et al. 2023; C. M. Casey et al. 2024), it is essential to fit the observed data with SPS models. To do this, fitting codes employ a statistical inference framework to determine which model spectral energy distributions (SEDs) most closely match the observed data. Models with emission lines allow full spectral fitting of data with fitting codes such as

★ E-mail: sophie.newman@port.ac.uk

CIGALE (D. Burgarella, V. Buat & J. Iglesias-Páramo 2005; M. Boquien et al. 2019), PPXF (M. Cappellari & E. Emsellem 2004; M. Cappellari 2017), BEAGLE (J. Chevallard & S. Charlot 2016), BAGPIPES (A. C. Carnall et al. 2018), and PROSPECTOR (B. D. Johnson et al. 2021), rather than masking the emission lines in the fitting process which is necessary with codes such as FIREFLY (D. M. Wilkinson et al. 2017) and STARLIGHT (R. Cid Fernandes et al. 2005, 2007).

SPS models are also crucial for analysing the results of hydrodynamical simulations, that track the formation and evolution of galaxies by modeling the physical processes governing gas dynamics, star formation, and feedback mechanisms. SPS models are used to translate simulation outputs, such as the ages and metallicities of star particles, into observable quantities like spectra, photometry and colours. Several studies have utilized SPS models to generate observables from cosmological simulations, including hydrodynamical and semi-analytic models (e.g. K. Park et al. 2015; G. F. Snyder et al. 2015; P. Torrey et al. 2015; C. Laigle et al. 2019; L. Nanni et al. 2022), including for *JWST* at high redshift (e.g. J.-P. Paardekooper, S. Khochfar & C. V. Dalla 2012; P. Dayal et al. 2013; H. Xu et al. 2016; D. Ceverino, S. C. O. Glover & R. S. Klessen 2017; W. I. Cowley et al. 2017; X. Ma et al. 2018; J. Rosdahl et al. 2018; L. Y. A. Yung et al. 2018). To predict the observable properties of young stars in simulations, it is important to couple simulations with SPS models featuring nebular emission (e.g. L. J. Kewley et al. 2013; J. W. Trayford et al. 2015; K. S. S. Barrow et al. 2017; S. Kaviraj et al. 2017; X. Shen et al. 2020; M. Vogelsberger et al. 2020; H. Katz et al. 2023a)

In this study, we introduce the Maraston models with nebular emission for the first time. A key feature of the Maraston models for young galaxies is their use of the Geneva stellar tracks (G. Schaller et al. 1992a; G. Meynet et al. 1994) for young ($t$ < 30 Myr) stellar populations. These tracks incorporate a number of different evolutionary assumptions, such as those regarding core-corrected overshooting and mass-loss. In this work, we use an updated version of the Maraston models incorporating the latest Geneva tracks (see Section 2), which allows us to assess the impact on the predicted SED, including nebular line and continuum emission, of parameters such as rotation and the effective temperature during the Wolf–Rayet (WR) phase.

Only a few existing SPS codes explicitly incorporate nebular emission in their standard predictions, such as PEGASE (M. Fioc & B. Rocca-Volmerange 1997) and STARBURST99 (C. Leitherer et al. 1999), though STARBURST99 includes only the nebular continuum. Both codes calculate this continuum using coefficients for hydrogen and helium from G. J. Ferland (1980), which characterize free-bound and bremsstrahlung (free–free) emission. These codes then use these theoretically derived coefficients to model the total continuous energy emitted across different wavelengths by the nebula. PEGASE also includes empirical line ratios obtained from local starburst galaxies.

The alternative, and more widely implemented, method for modelling emission lines is to use pure stellar SPS models as the incident ionizing source in a photoionization code. A number of works couple SPS models with photoionization codes in this way, including A. Feltre, S. Charlot & J. Gutkin 2016, who use the BC03 model, N. Byler et al. (2017) who use the FSPS model (C. Conroy et al. 2009a), L. Xiao, E. R. Stanway & J. J. Eldridge 2018 who use BPASS, B. Groves et al. (2008) who use STARBURST99, and M. Lecroq et al. 2024 who use the BC03 model including binary star processes.

Some of the most widely used photoionization codes are MAPPINGS-III (R. S. Sutherland & M. A. Dopita 1993; M. A. Dopita & R. S. Sutherland 1996; B. A. Groves, M. A. Dopita & R. S. Sutherland 2004) and CLOUDY (G. J. Ferland et al. 1998, 2017; M. Chatzikos et al. 2023; C. M. Gunasekera et al. 2023). In this work, we use CLOUDY due its integration in SYNTHESIZER[1]. (C. C. Lovell et al. 2025; W. J. Roper et al. 2025), a PYTHON package that enables the generation of synthetic spectra for cosmological and astrophysical applications. Once the user has provided the conditions of the gas cloud and the source of incident ionizing radiation (the SPS model chosen), CLOUDY simulates these physical conditions and predicts the resultant spectrum of the diffuse emission.

The paper is structured as follows. In Section 2, we describe the latest version of the Maraston SPS models that we use as our ionizing source in CLOUDY and its properties, such as the time evolution of the ionizing photon production rate. In Section 3, we describe the parameters and assumptions made in our photoionization modelling, including how we use SYNTHESIZER to generate grids of models. In Section 4, we then present our results, showing how our spectra with emission lines vary with age, metallicity, and ionization parameter. We also explore various diagnostic diagrams, and compare to other SPS models. Finally, we present our conclusions in Section 5.

## 2 INPUT THEORETICAL SPECTRA

Each SPS model begins with the time evolution of the spectral energy distribution (SED) of a simple stellar population (SSP), a coeval group of stars of the same metallicity. The spectrum of the SSP is obtained by summing the luminosity contribution from stars of different masses and in different evolutionary phases convolved with the assumed IMF.

Our study utilizes the Maraston stellar population models, which employ the Geneva tracks (G. Schaller et al. 1992a; G. Meynet et al. 1994) for stellar ages less than 30 Myr, and uses the fuel consumption theorem for the post main sequence (PMS) phases. In the fuel consumption theorem, the integration variable adopted for the phases of the post-main sequence is fuel, i.e. the product of luminosity and lifetime (A. Buzzoni 1989; C. Maraston 1998, 2005). On the other hand, in the commonly used isochrone synthesis approach the integral is performed using mass as the evolutionary variable (see G. Bruzual & S. Charlot 2003).

The Maraston models use the BaSeL stellar library (T. Lejeune, F. Cuisinier & R. Buser 1997, which was obtained by merging the theoretical Kurucz library (R. L. Kurucz 1979) and revisions) of model atmospheres with atmospheres for cooler stars. In particular, we use the M13 model, an update to the models developed by C. Maraston (2005), which apply the fuel consumption theorem to compute the energetics of the thermally pulsing asymptotic giant branch (TP-AGB) phase. The semi-empirical approach of Maraston uses Magellanic Cloud clusters to fix the energetics and colours for the population models featuring TP-AGB stars. This calibration is important as the ages assigned to the clusters determine when the TP-AGB phase occurs, and the integrated photometry of the clusters establishes the colours of the models.

Compared to the original models by C. Maraston (1998, 2005), the onset age of the TP-AGB phase in the newer M13 models is adjusted to match new calibrations using 43 Magellanic Cloud clusters as calculated and discussed in N. E. D. Noël et al. (2013). The fuel consumption during the TP-AGB phase has been set to zero up to an age of 0.6 Gyr, and the fuel consumption in that phase is also lower compared to the previous models for older ages. The TP-AGB contribution in the M13 models has been recently confirmed by the first detection of TP-AGB features in galaxies (S. Lu et al. 2024) at redshifts of $z \sim 1$–$2$.

[1] https://github.com/synthesizer-project/synthesizer

**Table 1.** Summary of differences between the Maraston stellar population models considered in this work.

| M05 | M13 | M24 |
|---|---|---|
| Uses the Geneva tracks for stellar ages less than 30 Myr, and then employs the fuel consumption theorem for the post-main-sequence (PMS) phases. Uses the BaSeL spectral library. | An update to the M05 model with new calibrations using 43 Magellanic Cloud clusters. | The same as the M13 model but with updated Geneva tracks with four variants:<br>⋆ No stellar rotation, corrected temperatures<br>⋆ Stellar rotation, corrected temperatures<br>⋆ No stellar rotation, uncorrected temperatures<br>⋆ Stellar rotation, uncorrected temperatures |

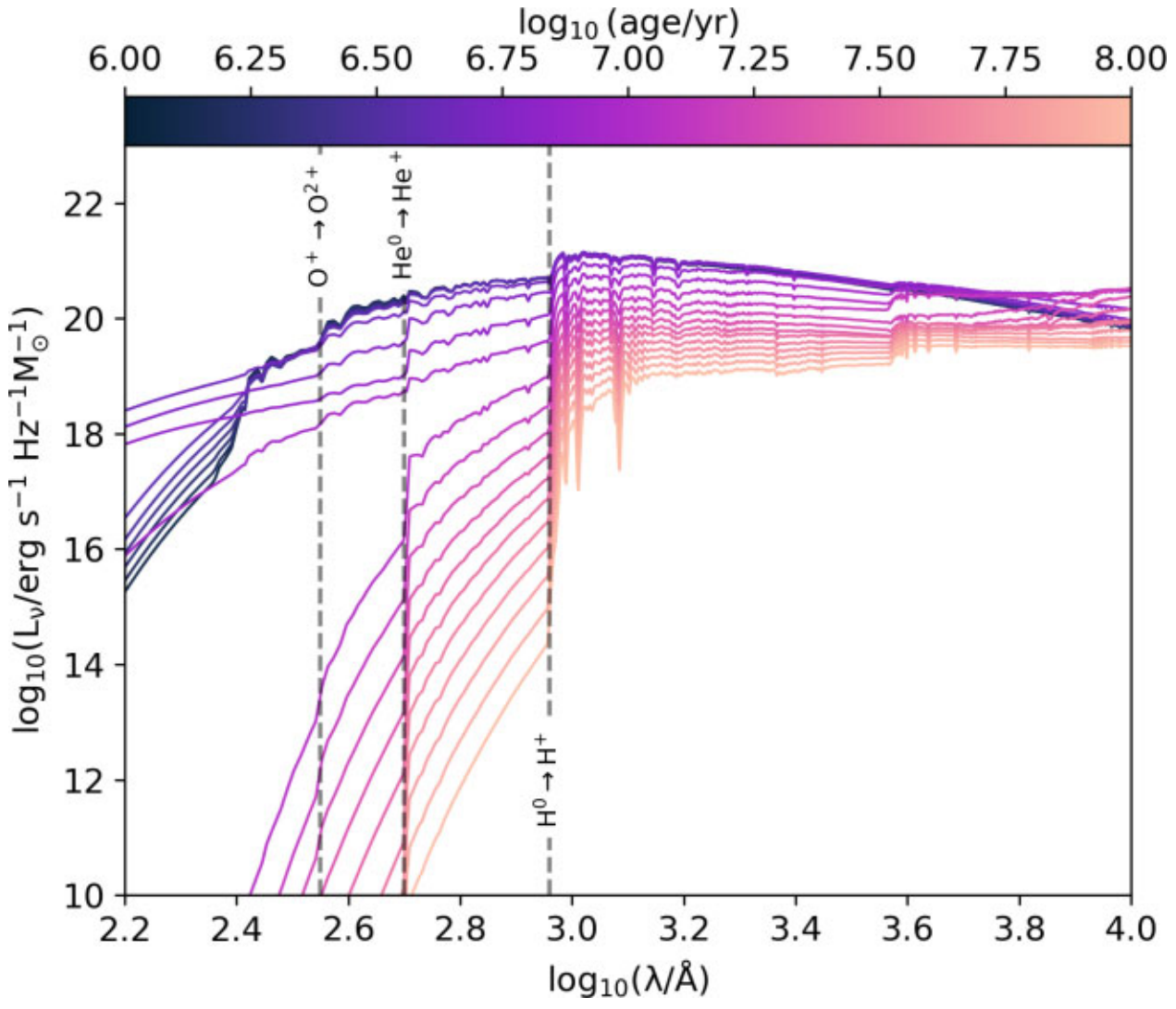

**Figure 1.** Our chosen M24 ionizing spectra for solar metallicity with ages varying from $10^6$ to $10^8$ yr, shown with dashed lines at the wavelengths associated with the energies required to ionize $H^0$ to $H^+$, $He^0$ to $He^+$, and $O^+$ to $O^{2+}$.

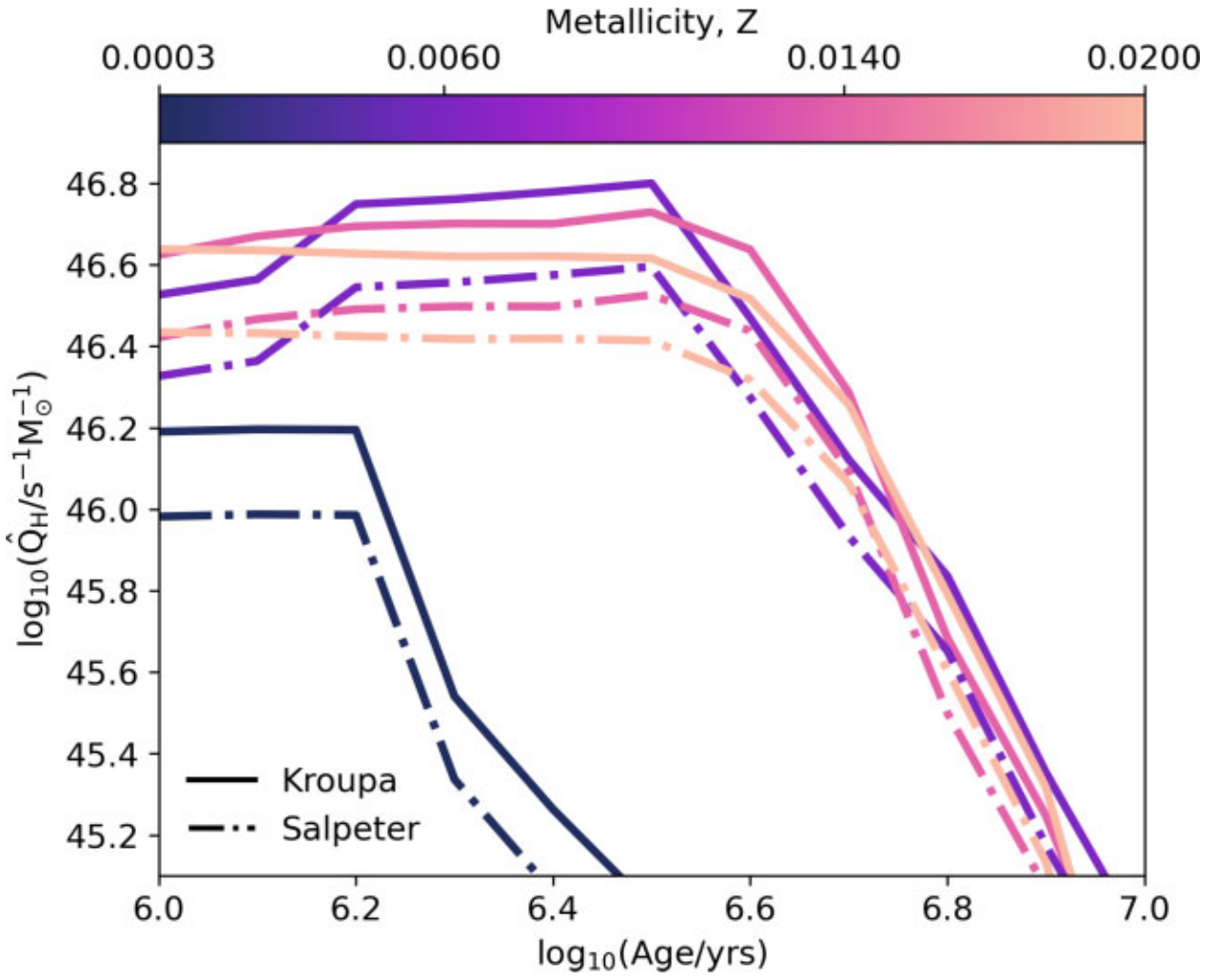

**Figure 2.** The specific hydrogen ionizing photon luminosity of the M24 simple stellar population model as a function of age with varying metallicity values ($Z = 0.0003$–$0.02$) and a Salpeter or Kroupa IMF.

In this paper we present an updated version of these models (referred to as M24 hereafter) as a source of ionizing radiation. The M24 model is the same as the M13 model but with updated Geneva tracks (S. Ekström et al. 2012; C. Georgy et al. 2013; J. H. Groh et al. 2019; P. Eggenberger et al. 2021; N. Yusof et al. 2022) that account for stellar rotation for ages $< 100$ Myr. The Geneva tracks were computed for both no rotation and rotation, with the rotating models having an initial equatorial rotational velocity of $v_{\rm init} = 0.4\, v_{\rm crit}$, where

$$v_{\rm crit} = \sqrt{\frac{2}{3}\frac{GM}{R_{\rm pol,crit}}}\,, \qquad (1)$$

and $R_{\rm pol,crit}$ is the polar radius at $v_{\rm crit}$. Models are also calculated using corrected or uncorrected effective temperatures for the non-negligible optical winds of WR stars, the details of which can be found in Section 2.7 of G. Schaller et al. (1992b). Therefore, we have four variants of the new M24 models and we summarize these as well as the M05 and M13 models in Table 1. In this work, we assume the rotating model with uncorrected temperatures as our fiducial M24 model, since it produces the highest number of ionizing photons.

The M24 models have been computed for 21 ages between $10^6$ and $10^8$ yr since the aim of this work is to study the properties of young star-forming populations. The models are computed for five metallicity values, $Z = [0.0003, 0.002, 0.006, 0.014, 0.02]$ where $Z = 0.014$ is assumed to be the solar metallicity value. We make the full set of models available online[2].

## 2.1 Characteristics of the model ionizing spectra

Since it is important to consider the properties of the ionizing radiation that is fed into CLOUDY, in this Section, we will discuss how these properties of our chosen Maraston model (M24) vary with age and metallicity, and how they compare with other available choices of SPS model and IMF.

In Fig. 1, we show the M24 spectra plotted for each age for a fixed solar metallicity. Also shown are dashed lines at the wavelengths associated with the energies required to ionize $H^0$ to $H^+$, $He^0$ to $He^+$, and $O^+$ to $O^{2+}$. Here, we can see that the gradient of the extreme ultraviolet (EUV) slope, defined between the ionization energies of $He^0$ and $H^0$, generally increases with age; older populations have less high-energy photons that can ionize a model gas cloud. Post-AGB hot stellar phases can boost ionizing fluxes in older populations (e.g. R. Yan & M. R. Blanton 2012); while post-AGB phases are included in the Maraston models (see F. Belfiore et al. 2016), we do not discuss them further in this paper, and focus on young stellar populations.

We plot the hydrogen ionizing photon production rate for each of these models as a function of age and metallicity in Fig. 2. As expected, the youngest models (age $\lesssim 4$ Myr) produce the highest number of ionizing photons per second, with a noticeable increase between 1 and 4 Myr due to the onset of the Wolf–Rayet (WR) phase. Metallicity also plays a significant role: the intermediate-low metallicity model ($Z = 0.006$) produces the most ionizing photons at fixed age and IMF – for example, around 3 Myr – owing to reduced line blanketing and hotter stellar temperatures. However, this trend does not continue monotonically toward lower metallicities. The extremely metal-poor model ($Z = 0.0003$) shows a lower ionizing photon production rate, likely due to effects such as weaker stellar winds. Thus, while lower metallicity generally enhances ionizing output, this effect peaks around $Z = 0.006$ and is less pronounced or even reversed at very low metallicities. Since both nebular continuum and line emission strength are wholly dependent on the ionizing photon production rate, the differences in ionizing photon production rates for different SPS models will be discussed in Section 4.4. We checked the effect of varying the mass-loss rate in the input stellar tracks for the M13 models and found it has little consequence on the ionizing photon production rate, as discussed in Appendix A.

[2] `https://sophie-newman.github.io/cloudy-maraston.html`

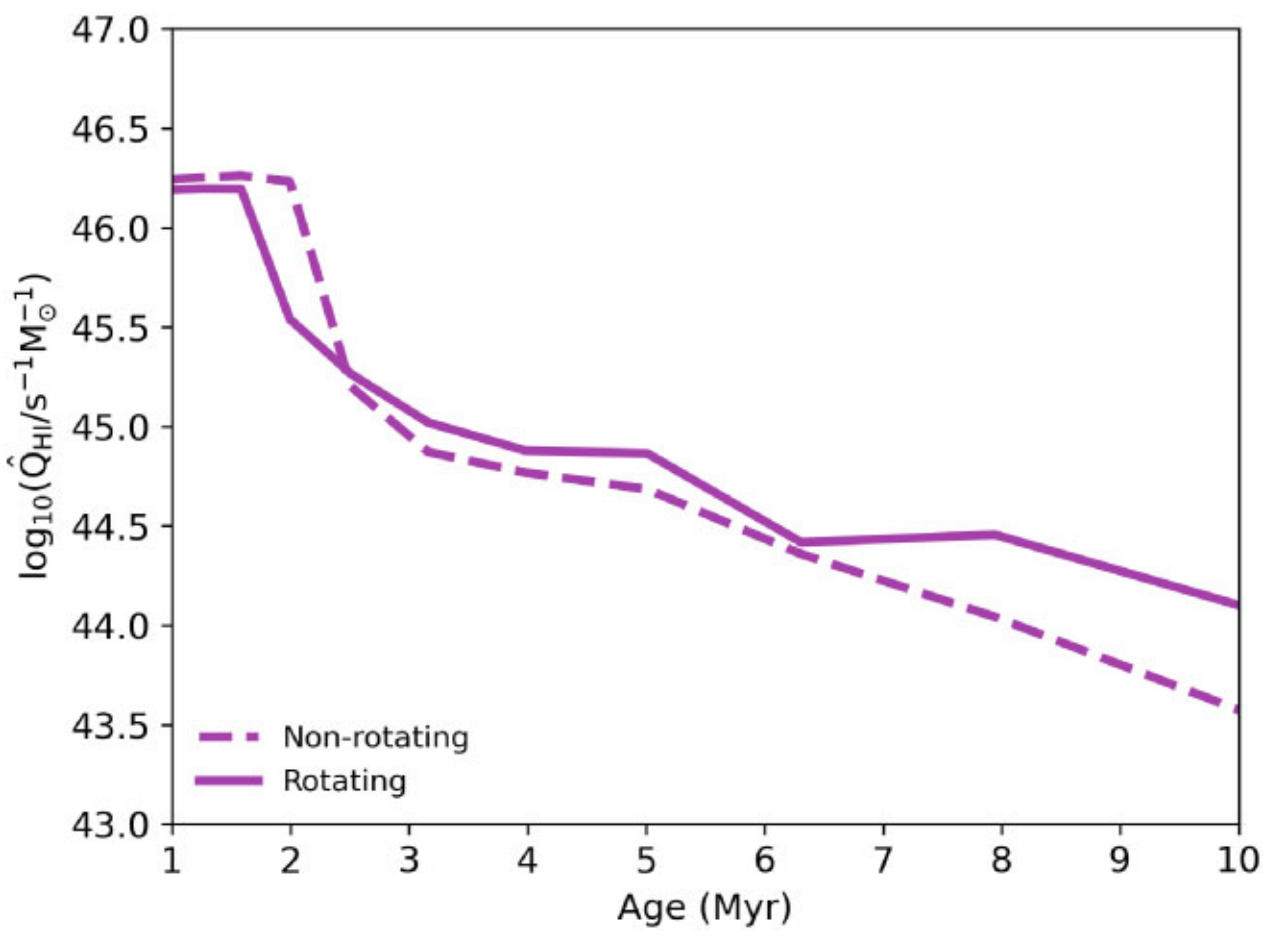

**Figure 3.** The specific ionizing photon luminosity of the M24 model as a function of age for solar metallicity ($Z = 0.014$) for different rotations.

In Fig. 3, we also show the ionizing photon production rate for the M24 model with and without rotation. We find that including rotation can boost the number of hydrogen ionizing photons being produced which we will see again in Section 4.4. However, we do not observe the ionizing photon production rate changing between the models with corrected effective temperatures during the WR phase or those without.

### 2.2 Varying IMF

The luminosity of a stellar population is strongly dependant on the form of the stellar initial mass function (IMF; C. Conroy et al. 2009b; N. Bastian, K. R. Covey & M. R. Meyer 2010). Top-heavy IMFs, where a larger fraction of the total mass is in high-mass stars, particularly affect young populations where massive stars dominate. Commonly used forms of the IMF include a power law, such as E. E. Salpeter (1955), a multisegment power law such as P. Kroupa (2001), or G. Chabrier (2003), which is very similar to the Kroupa IMF since it resembles a two-part power law. In this paper we consider the E. E. Salpeter (1955) and P. Kroupa (2001) parametrizations that have been implemented for the M24 models. However, models for any arbitrary IMF can be calculated upon request.

In Fig. 2, we see the effect of varying the IMF on the ionizing photon luminosity as a function of age. At an age of $10^6$ yr and at fixed metallicity, the ionizing photon luminosity of the broken Kroupa IMF is 1.6 times higher than that of the Salpeter IMF. This is because the Kroupa IMF is top-heavy relative to Salpeter. Due to the steep luminosity dependence on stellar mass, the higher abundance of high temperature massive stars featuring a Kroupa IMF results in more hydrogen ionizing photons and more nebular emission. Higher metallicity models ($Z = 0.014$–$0.02$) produce the most ionizing photons up to around 1.5 Myr, after which lower metallicity models then begin to dominate. In the modelling of UV to radio SEDs, I. A. Obi et al. (2017) also observed this trend with metallicity for varying upper mass limits of the initial mass function of $M_{up} = 120\,M_\odot$ and $50\,M_\odot$ but not for the lower $M_{up} = 40\,M_\odot$. This demonstrates how the most massive stars affect the number of ionizing photons produced and the importance of considering the dependence of the IMF on the ionizing photon production rate. In Section 4, we will assume a Kroupa IMF.

## 3 METHODS

In this section, we discuss the parameters within the emission model and our means of marrying it with the Maraston model with the use of SYNTHESIZER.

### 3.1 `Cloudy` modelling

To calculate and add nebular emission to the M24 models, we use version 23.01 of CLOUDY[3], described in M. Chatzikos et al. (2023) and C. M. Gunasekera et al. (2023). CLOUDY is a photoionization code developed to simulate the physical conditions within a gas cloud and predict the resulting emission spectrum. The CLOUDY user must specify the external radiation source to ionize the gas cloud, as well as the physical properties of the cloud. In this section, we outline our choices for the cloud's geometry and chemical composition, as well as how we quantify the intensity of the ionizing radiation source.

Like many others in the literature (e.g. S. Charlot & M. Longhetti 2001; A. Feltre et al. 2016; J. Gutkin, S. Charlot & G. Bruzual 2016; M. Hirschmann et al. 2017), we choose to characterize our modelling by the dimensionless ionization parameter, $U$, and the hydrogen density (cm$^{-3}$), $n_H$. The ionization parameter $U$ is defined as the ratio of hydrogen ionizing photons to total hydrogen density:

$$U(R) = \frac{Q_H}{4\pi R^2 \cdot n_H \cdot c} \tag{2}$$

where $Q_H$ is the total number of photons emitted per second that are capable of ionizing hydrogen, $c$ is the speed of light, and $R$ is the characteristic radius at which $U$ is evaluated. The quantity $Q_H$ reflects the intensity of the ionizing spectrum and is given by

$$Q_H = \frac{1}{hc}\int_0^{\lambda_0 \leq 912\text{Å}} \lambda L_\lambda \mathrm{d}\lambda \ , \tag{3}$$

where $L_\lambda$ is the luminosity of the SPS model at a specific wavelength, and 912Å is the Lyman-limit, the wavelength corresponding to the ionization energy of hydrogen.

A common choice for the radius $R$ is the Strömgren radius, $R_S$, which defines the extent of the Strömgren sphere – the region in which all hydrogen atoms are ionized. This radius is determined by balancing the rates of ionization and recombination, and is therefore

[3] https://gitlab.nublado.org/cloudy/cloudy

given by

$$R_S^3 = \frac{3Q_H}{4\pi n_H^2 \epsilon \alpha_B} \quad , \tag{4}$$

with $\epsilon$ denoting the volume-filling factor of the gas, and $\alpha_B$ the Case B hydrogen recombination coefficient (D. E. Osterbrock 1989). For typical electron temperatures within H II regions of $T_e \approx 10^4$ K, we adopt a value of $\alpha_B \approx 2.6 \times 10^{-13}$ cm$^3$ s$^{-1}$. However, since $\alpha_B$ depends on $T_e$, we note that variations in the electron temperature can influence the resulting value of $U$.

In CLOUDY, the strength of the ionizing source can be provided in terms of the ionization parameter defined at the inner edge of the cloud, $R_{inner}$. Alternatively, rather than evaluating the ionization parameter at a specific radius, we can use the *volume averaged* ionization parameter, $\langle U \rangle$, which gives the average of the ionization parameter throughout the ionized region, and is the definition we will adopt throughout the rest of the text, and refer to simply as $U$ beyond this section.

The geometry of the ionized region is essentially determined by the ratio of the radius of the inner edge of the cloud and the Strömgren radius. We assume a spherical geometry, where $R_{inner} << R_S$, the ionization region has a thickness approaching $R_S$, and the ionization parameter depends strongly on $R$. This gives the following for the volume averaged ionization parameter with $dV = 4\pi R^2 dR$,

$$\langle U \rangle = \frac{1}{V} \int_0^{R_S} U(R)\, dV = \frac{3Q_H}{4\pi n_H c R_S^2} \tag{5}$$

$$= \frac{\alpha_B^{2/3}}{c} \left( \frac{3Q_H \epsilon^2 n_H}{4\pi} \right)^{1/3} \quad . \tag{6}$$

For a fixed $\langle U \rangle$ and $Q_H$, the geometry of the region is encoded in the $\epsilon^2 n_H$ term. The utility of $\langle U \rangle$ is that it helps reduce the dimensionality of our input grid substantially; any combination of $Q$, $\epsilon^2$, and $n_H$ that gives the same $\langle U \rangle$ will produce the same nebular spectrum (assuming a fixed ionizing spectrum shape and fixed metallicity and abundance pattern in the cloud).

We begin our CLOUDY calculations at $R_{inner} = 0.01$ pc, which is valid to use as the lower integration limit in equation (5) since this is a negligible radius compared to $R_S$. Therefore the only values that we choose to fix or vary are $\langle U \rangle$, $Q_H$, and $n_H$. We explore two different approaches:

(i) **Fixed $\langle U \rangle$ approach:** In this method, we fix either the volume-averaged ionization parameter $\langle U \rangle$ or the ionizing photon rate $Q_H$ to achieve the target $\langle U \rangle$, with a fixed hydrogen density $n_H$. Since the $Q_H$ predicted by single stellar population models for a 1 $M_\odot$ population is often insufficient to produce the desired ionization parameter, we follow the normalization strategy from N. Byler et al. (2017). Specifically, we compute the required ionizing photon rate per unit solar mass ($\hat{Q}_H$) and derive the corresponding normalization mass $\hat{M}$ that scales $Q_H$ to the target $\langle U \rangle$. This $\hat{M}$ is then used to renormalize the resulting emission-line predictions from CLOUDY. We explore a grid of models spanning $\langle U \rangle = [10^{-4}, 10^{-3}, 10^{-2}, 10^{-1}]$ and $n_H = [10, 10^2, 10^3, 10^4, 10^5]$ cm$^{-3}$.

(ii) **Free $\langle U \rangle$ approach:** In this method, we allow the ionization parameter to vary naturally based on the properties of the input stellar population, without adjusting $Q_H$ to match a fixed $\langle U \rangle$. Instead, we fix the normalization mass $\hat{M}_{ref}$ based on a reference model – specifically, an SSP with an age of 1 Myr and metallicity $Z = 10^{-2}$ that produces $\langle U \rangle = 10^{-2}$ at $n_H = 100$ cm$^{-3}$. This $\hat{M}_{ref}$ is then used across all models, allowing $\langle U \rangle$ to vary in response to changes in the ionizing spectrum as a function of age and metallicity. This approach captures the evolution of $\langle U \rangle$ across different stellar populations without manually tuning $Q_H$.

The values of ionization parameter that are varied in the first approach are consistent with those derived from local starburst galaxies (e.g. M. A. Dopita et al. 2000; M. D. Thornley et al. 2000; J. R. Rigby & G. H. Rieke 2004; J. Moustakas, J. Kennicutt & C. A. Tremonti 2006; J. Moustakas et al. 2010), as well as those inferred from high-redshift galaxies observed by *JWST* (e.g. A. Calabro et al. 2024; M. Castellano et al. 2024; T. Y.-Y. Hsiao et al. 2024; W. Hu et al. 2024; M. Llerena et al. 2024). They can be inferred from observations using line ratios such as [O III]$\lambda$5007/[O II]$\lambda$3726,3729 (O32); these line ratios will be analysed in Section 4. For a fixed geometry and hydrogen density, the highest ionization parameters considered in the first approach represent low metallicity stars that produce more ionizing photons for a given stellar mass. For the considered hydrogen densities, our chosen parameters cover most of the range of electron densities (which can be estimated as $\sim n_H$) observed locally (e.g. X. Liu et al. 2008; L. K. Hunt & H. Hirashita 2009; T. M. Hughes et al. 2017; R. L. Davies et al. 2021) and at higher redshifts (e.g. R. L. Sanders et al. 2016; M. W. Topping et al. 2020; N. A. Reddy et al. 2023; A. Mizener et al. 2024).

To set up the CLOUDY models, commands such as `sphere`, `radius`, `hden`, and `Q(H)` are used to set up the geometry, inner radius, hydrogen density, and $Q_H$. We use the Galactic Concordance model, introduced by D. C. Nicholls et al. 2017, based on extensive Milky Way stellar abundance data. This provides a reference abundance pattern (see their table 2) together with metallicity-dependent scaling relations for a broad set of chemical elements. For most elements, abundances are scaled relative to oxygen through a simple linear relation, which flattens to plateaus at both low and high oxygen-to-hydrogen ratios (O/H). Carbon and nitrogen are treated with more detailed prescriptions to account for the unique and complex nucleosynthetic processes – such as metallicity dependence, stellar mass effects, and secondary production – that govern their abundances (F. Matteucci 1986; A. Maeder 1992; F. Vincenzo et al. 2016). Specifically, their abundances are defined by:

$$\log_{10}\left(\frac{N}{H}\right) = \log_{10}\left(10^{-1.732} + 10^{(O/H + 2.19)}\right) \tag{7}$$

$$\log_{10}\left(\frac{C}{H}\right) = \log_{10}\left(10^{-0.8} + 10^{(O/H + 2.72)}\right) \tag{8}$$

For the lightest elements (e.g. H, He, Li), no metallicity scaling is applied in this framework. Their abundances are fixed at the adopted reference values, consistent with solar composition. These reference abundances and those calculated for each metallicity case considered in this work are presented in Appendix B.

The gas-phase abundances of elements are often lower than the total abundances due to the incorporation of elements into dust grains. Depletion models provide a framework to relate the gas-phase, dust-phase, and total abundances of elements. In general, the gas-phase abundance of an element $X$ can be expressed as

$$\left(\frac{X}{H}\right)_{gas,dep} = D_X \times \left(\frac{X}{H}\right)_{total} \tag{9}$$

where $D_X$ is the depletion factor. The corresponding abundance locked in dust is then

$$\left(\frac{X}{H}\right)_{dust} = (1 - D_X) \times \left(\frac{X}{H}\right)_{total} \quad . \tag{10}$$

Here, we adopt the E. B. Jenkins 2009 depletion pattern as implemented in CLOUDY by C. M. Gunasekera et al. 2022, which extends the original model with additional elements. In this model, the depletion factor $D_X$ for each element is parameterized as

$$D_X = 10^{B_X + A_X(F_* - z_X)} \tag{11}$$

where $A_X$, $B_X$, and $z_X$ are empirically fitted coefficients and $F_*$ is a scaling factor describing the overall depletion strength along a line of sight. For the purposes of this work, we assume $F_* = 0.5$ to represent a moderate level of depletion. The resulting depletion factors are presented in Appendix B. Elements with higher $D_X$ are less depleted, whereas those with lower $D_X$ are more strongly incorporated into dust grains. Once calculated, the depleted abundances are then provided to CLOUDY with the `element abundance` command, resulting in dust-to-metal ratios of around $\sim 0.4$. In addition, we include graphite, silicate, and polycyclic aromatic hydrocarbon (PAH) grains. The graphite and silicate grains adopt the Orion nebula grain size distribution (dominated by large grains) and elemental abundances implemented in Cloudy, invoked through the commands `grains Orion graphite` and `grains Orion silicate`, respectively. PAH grains are included by assuming that 1 per cent of the total carbon is locked in PAHs, using the command `grains PAH`. The dust scaling is set such that the total elemental mass is conserved, ensuring that the adopted depletions correspond to the mass contained in dust.

We choose to stop our calculations when the free electron density falls to 0.01 of the hydrogen density using the `efrac -2` command. The CMB can become a significant source of heating for dust and gas at high redshift ($z > 5$) and affect the observed (sub)millimeter nebula emission (F. Combes, R. Maoli & A. Omont 1999; E. da Cunha et al. 2013). Since we are not focusing on this wavelength regime, we do not consider CMB heating in this study.

We assume that the total nebular metallicity – which includes all contributions from metals depleted onto dust – and the abundance pattern are consistent with those of the stellar model. This is a simplification since, in the absence of gas inflows, the nebular metallicity in star-forming regions in the local universe is greater than the stellar metallicity (e.g. A. Gallazzi et al. 2005; J. Lian et al. 2018a; A. Fraser-McKelvie et al. 2022). Nebular metallicity has been found to increase with decreasing redshift from $z = 3$ to $z = 0.1$ (J. Lian, D. Thomas & C. Maraston 2018b; N. M. Förster Schreiber & S. Wuyts 2020), and *JWST* observations from $z = 3$ to $z = 9$ have demonstrated a continued gradual evolution towards higher total nebular metallicities (M. Curti et al. 2023; T. Morishita et al. 2024) at lower redshifts. We expect that stellar metallicities will be similar to nebular metallicities at higher redshift, since recently formed stars will have been created from a very similar ISM to that present in the new H II regions. We focus on younger stellar populations in this work where the gas and stellar metallicities are likely to be closer in value.

In future work by Wilkins et al., the effects of and reasoning behind several assumptions made in this section will be examined, including the use of plane-parallel rather than spherical geometry, a fixed ionization parameter and various abundance patterns. The impact of depletion, and dust grains more generally, is explored in Vijayan et al. (in preparation).

### 3.2 Integration of nebular emission with SYNTHESIZER

We use SYNTHESIZER (C. C. Lovell et al. 2025; W. J. Roper et al. 2025), a code for generating multiwavelength emission from a wide range of astrophysical models with varying levels of complexity and precision. At one end of the spectrum, SYNTHESIZER can create simple toy models that describe a galaxy through analytic forms. At the other end, it can ingest data from high-resolution, isolated galaxy simulations, incorporating tens of thousands of discrete elements representing the galaxy's stellar and gas properties. SYNTHESIZER is designed to be modular, flexible, and efficient, and its framework can be applied to a variety of tasks, not limited to producing observables from cosmological simulations.

Specifically, we utilize the GRID-GENERATION[4] package, which provides scripts and tools for generating grids of emission from various astrophysical sources. Grids are a key component of SYNTHESIZER; at their most basic level, they describe emission as a function of specific parameters, typically the age and metallicity of a stellar population, with the emission derived from a stellar population synthesis (SPS) model. However, the parameters can be arbitrary, multidimensional, and the emission can represent any source. For instance, the source might be the emission from the narrow line region of an active galactic nucleus (AGN).

We employ the two main modules of GRID-GENERATION: one for generating incident grids (spectra produced by a model) and another for creating and running CLOUDY input files, which are then combined to form a SYNTHESIZER grid.

First, we use GRID-GENERATION to convert the original file containing the Maraston SPS model to a grid in the format that SYNTHESIZER uses. Then this grid is used to generate a set of input files, that contain the chosen modelling parameters, that can be processed by CLOUDY, with the increased value of $Q_{\rm H}$ described in Section 3.1. Once the code has been run through CLOUDY, GRID-GENERATION is then used to re-normalize the outputs to match the 1 $M_\odot$ incident SEDs that were first given. For more details for how SYNTHESZIER models photoionization, see Wilkins et al. (in preparation) and the `c23.01-sps.yaml` file[5] within GRID-GENERATION.

## 4 RESULTS

In Fig. 4, we show the rest-frame optical spectrum from a 1 Myr model with solar metallicity and an ionization parameter of $U = 10^{-2}$ processed with CLOUDY, with its different components shown. The components highlighted here are the nebular emission, nebular continuum, transmitted emission, incident emission, and total emission. These are defined using the following definitions:

Incident: the spectra that serve as an input to CLOUDY as discussed in Section 3.1. In the context of stellar population synthesis, we used the M24 models that are equivalent to the 'pure stellar' spectra.

Transmitted: the incident spectra that is transmitted through the gas in our photoionization modelling. The main difference between transmitted and incident is that the transmitted has little flux below the Lyman-limit since this has been absorbed by the gas.

Nebular: the nebular continuum and line emission predicted by the CLOUDY photoionization model.

Total: the sum of the transmitted and nebular emission.

In Fig. 5, we show the total emission for solar metallicity models with an ionization parameter of $U = 10^{-2}$ and ages varying between $10^6$ and $10^8$ yr. As expected, models with the youngest ages have the strongest line emission. We observe an increase in luminosity for models with an age around $10^7$ yr due to the emergence of the red supergiant phase. Additionally, the overall shape of the spectrum changes with age, including variations in the UV slope, which are mainly driven by nebular continuum emission. This will be discussed in the following subsection.

[4] https://github.com/synthesizer-project/grid-generation
[5] This file can be found here.

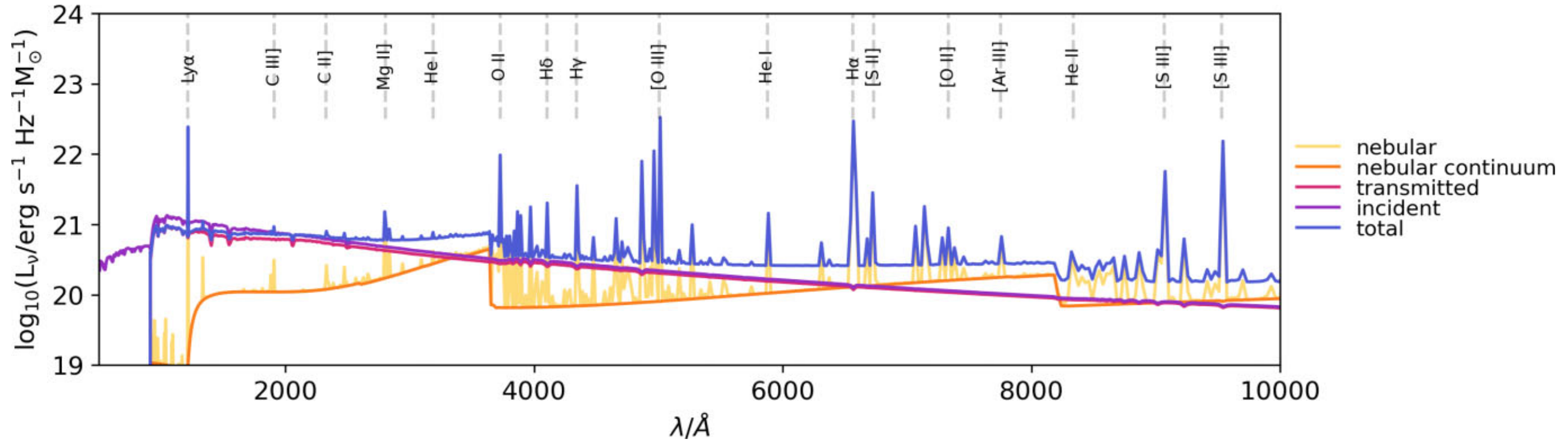

**Figure 4.** The components of a M24 spectrum with an age of 1 Myr, solar metallicity and a reference ionization parameter of $U = 10^{-2}$. Here, the incident component is the spectra that ionizes the cloud in our photoionization modelling, transmitted is the incident spectra that is transmitted through the cloud, nebular is the sum of the nebular continuum and line emission, and the total is the sum of the transmitted and nebular emission. Labelled are the positions of visible emission lines from optical to near-infrared wavelengths.

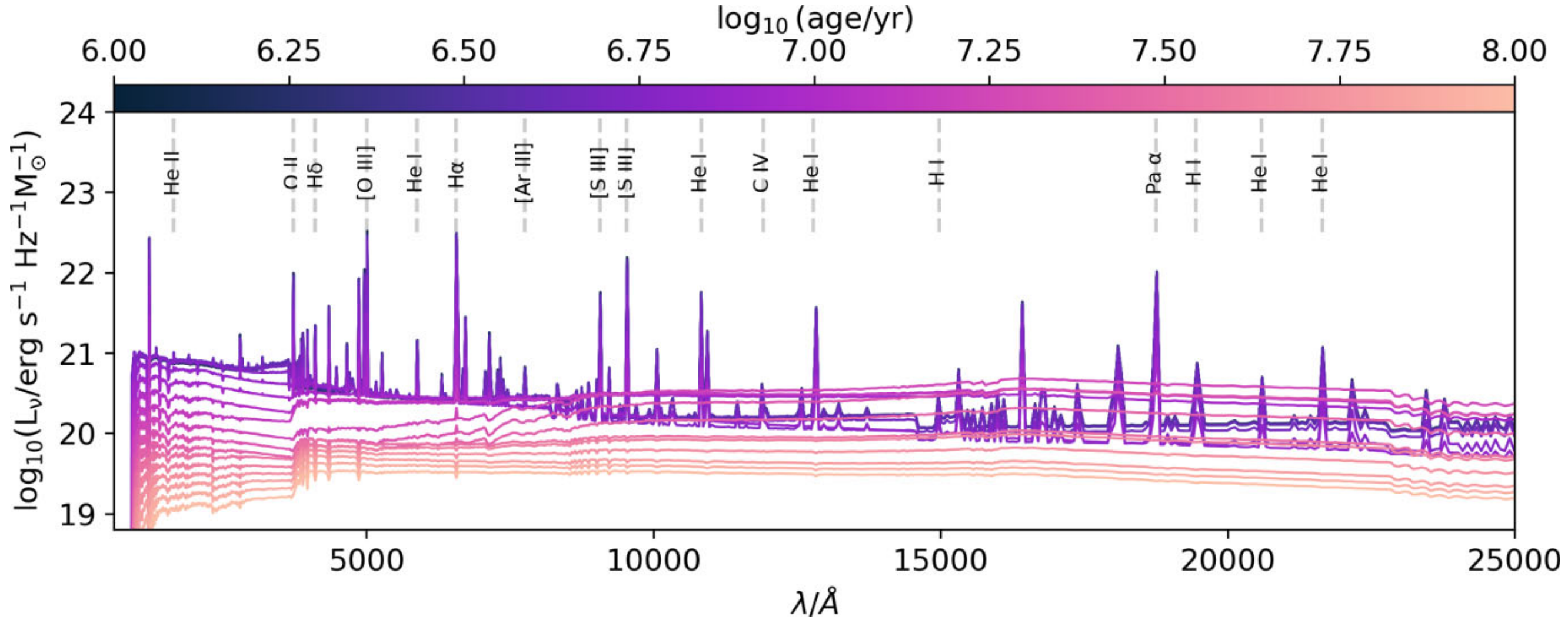

**Figure 5.** The total emission for the M24 solar metallicity models with a reference ionization parameter of $U = 10^{-2}$ and ages varying between $10^5$ and $10^7$ years. The positions of common emission lines, as well as a CO molecular transition, from optical to mid-infrared are indicated with dashed lines.

### 4.1 Nebular continuum

The nebular continuum represents the collective emission from various processes occurring within nebulae, including the free-bound (recombination) transition, free–free (Bremsstrahlung) transition, and two-photon emission. This continuum spans a broad range of wavelengths, from ultraviolet to infrared, and can contribute significantly to the observed spectrum. For example, A. E. Reines et al. (2010) discovered that the nebular continuum can contribute around 40 per cent of the total I-band ($\lambda \approx 8500$–$7600$ Å) flux in young ($< 5$ Myr) massive stellar clusters. In recent work, H. Katz et al. 2024 showed how nebular continuum can significantly increase the UV luminosity and lead to reddened UV slopes. Since the free–free continuum is relatively flat in the infrared and falls off steeply toward optical wavelengths, while the free-bound continuum dominates in the optical, it becomes an increasingly important contributor to the total nebular continuum at near-infrared wavelengths.

Describing the nebular continuum requires understanding the dependence on parameters such as metallicity, ionization parameter, and hydrogen density, whose effects we will explore in this Section. In Figs 6(a) and 6(b), for a model with a fixed age of 1 Myr, we see that the intensity of the nebular continuum is affected by both metallicity and ionization parameter. In particular, we observe that the models with the lowest ionization parameters and nebular metallicity have the highest nebular continuum, in agreement with similar modelling done by M. L. P. Gunawardhana et al. (2020). At the wavelengths corresponding to the limits ($n \to \infty$) of different series of Hydrogen (Lyman, Balmer, Paschen), we observe the distinctive sawtooth recombination edges due to the free–bound continuum (for more details see e.g. B. T. Draine 2011). We also observe a distinct bump at $\sim 1500$ Å, between the Lyman and Balmer discontinuities, that is also caused by the two-photon continuum.

Here, we also see that the ionization parameter scales the normalization of the nebular continuum up or down. This is because a higher ionization parameter leads to more frequent recombinations and since free-bound emission is a result of recombination, its intensity increases with the number of ionizing photons. Metallicity (which is inversely proportional to the electron temperature and therefore the modelled nebular temperature) appears to have a smaller effect on the nebular continuum however, with the higher metallicity models being associated with recombination edges of greater relative height and steepness. When dust grains are excluded from the CLOUDY modelling, the influence of metallicity and the ionization parameter on the continuum strength significantly decreases.

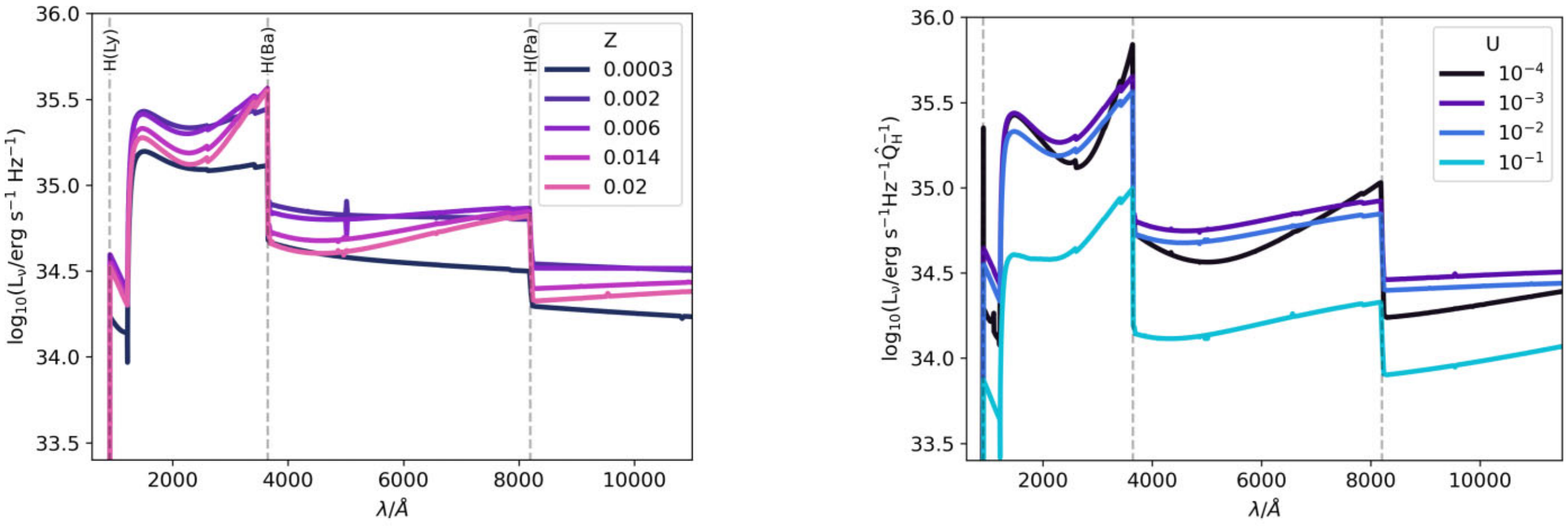

(a) Varying metallicity

(b) Varying ionization parameter

**Figure 6.** The nebular continuum modelled with CLOUDY for various metallicities $Z$ (from 0.0003 to 0.02) and ionization parameters $U$ (from $10^{-4}$ to $10^{-1}$) for a fixed age of 1 Myr. Labelled also are discontinuities at the wavelengths corresponding to the limits ($n \rightarrow \infty$) of different series of Hydrogen (Lyman, Balmer, and Paschen).

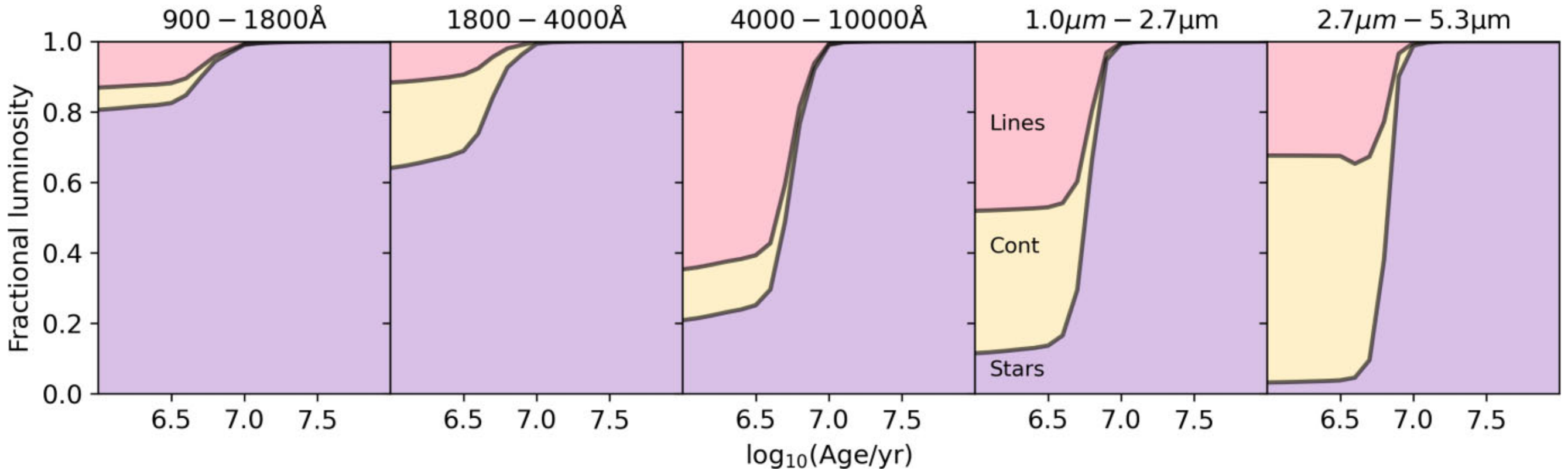

**Figure 7.** Fractional contribution to the total luminosity from stars, the nebular continuum, and line emission, for solar metallicity ($Z = 0.014$). Each panel shows the fractional contribution as a function of the model age for various wavelength bands: far-ultraviolet (900–1800 Å), near-ultraviolet (1800–4000 Å), optical (4000–10 000 Å), NIR (1.0–2.7 μm), and MIR (2.7–5.3 μm).

Shown in Fig. 7 is the fractional contribution to the total luminosity from the nebular continuum, as well as the contribution from stars and emission lines, in various wavelength ranges from the UV to the IR. We see, as expected, that the contribution from the nebular continuum increases approaching the near-IR band, and this is also seen in the total emission in Fig. 5. The stellar contribution is at its maximum in the far-ultraviolet and near-ultraviolet, whereas the contribution from emission lines to the total luminosity is at its maximum in the optical band. The increase in the stellar contribution between $10^6$–$10^7$ yr inversely correlates with the production rate of ionizing photons seen in Fig. 2.

The results for the nebular continuum contributions in different wavelength ranges are particularly interesting since observations by *JWST* have pointed towards the existence of ‘nebular-dominated galaxies’ where the UV continuum appears to be highly nebular dominated (e.g. A. J. Cameron et al. 2024; H. Katz et al. 2024). While we do find that the contribution from the continuum increases with wavelength like these studies observe, our models have lower continuum fractions. For example, at 1500 Å the BPASS model in H. Katz et al. (2024) has a total nebular (line + continuum) contribution of ∼ 40 per cent, higher than our contribution of ∼ 20 per cent. In the work by H. Katz et al. (2024), they find lower limits on the nebular contribution fractions of *JWST* data described in L. Barrufet et al. (2024), A. J. Cameron et al. (2024), A. Saxena et al. (2024), and K. Boyett et al. (2024) of 60–80 per cent. Many different explanations have been given for this observed nebular domination (A. J. Cameron et al. 2024; K. E. Heintz et al. 2024; Y. Li et al. 2024; D. Schaerer et al. 2024; S. Tacchella et al. 2024; C. Terp et al. 2024; H. Yanagisawa et al. 2024), with some studies suggesting that the observed sources are not in fact nebula dominated.

### 4.2 Evolution of emission lines

In Fig. 8, the relationships in line strength for various strong optical emission lines are displayed as a function of age and varying metallicities, ranging from 1 to 10 Myr. The lines shown are H$\beta\lambda$4861, [O III]$\lambda$5007, H$\alpha\lambda$6563, [N II]$\lambda$6583, and [S II]$\lambda$6716,6731. The [S II]$\lambda$6716,6731 doublet is used often to estimate electron densities (e.g. N. A. Reddy et al. 2023), while [N II]λ6583 and [O III] are commonly used to probe ISM conditions in the O3N2 diagram at $z \leq 2.5$ (A. E. Shapley et al. 2023). Generally, hydrogen lines are used to normalise out the strong line metallicity calibrations. In Fig. 9, we also vary the ionization parameter from $U = 10^{-1}$ to $U = 10^{-4}$ for a fixed model metallicity of $Z = 0.014$. Throughout this work, the wavelengths mentioned follow the same convention used by CLOUDY, specifically: vacuum wavelengths are used below 200 nm, and air wavelengths for above 200 nm.

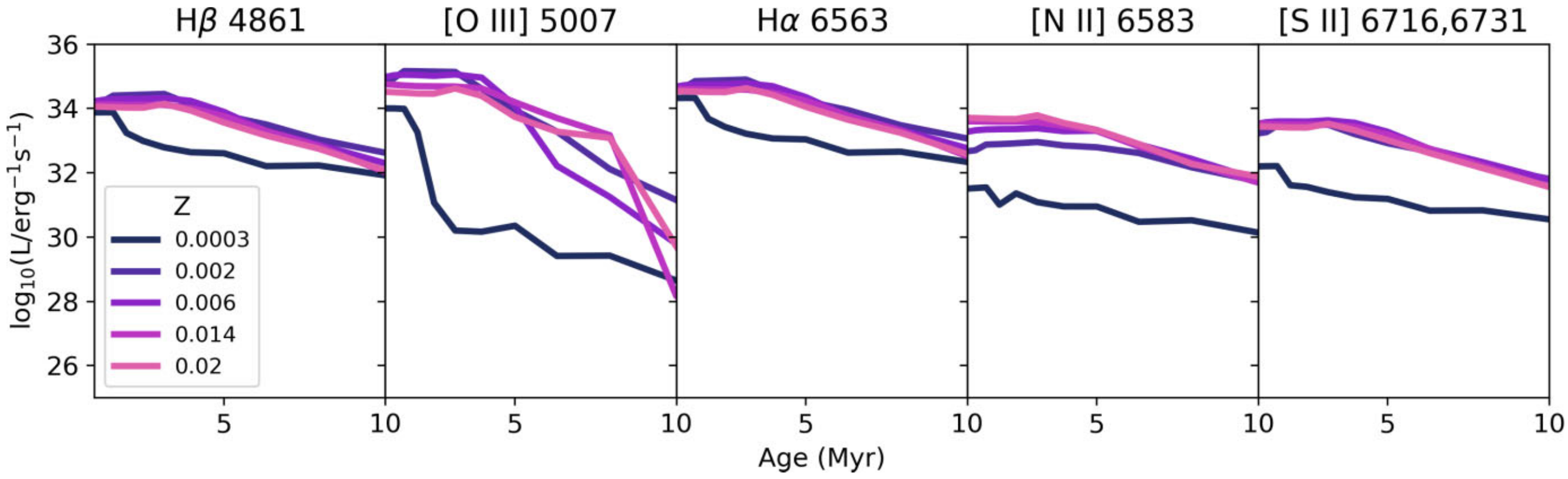

**Figure 8.** Emission strength for the lines Hβλ4861, [O III]λ5007, Hαλ6563, [N II]λ6583, and [S II]λ6716,6731 as a function of the age of the model for varying model metallicity and a fixed ionization parameter of $U = 10^{-2}$. Here the metallicity $Z$ varies from 0.0003 to 0.02.

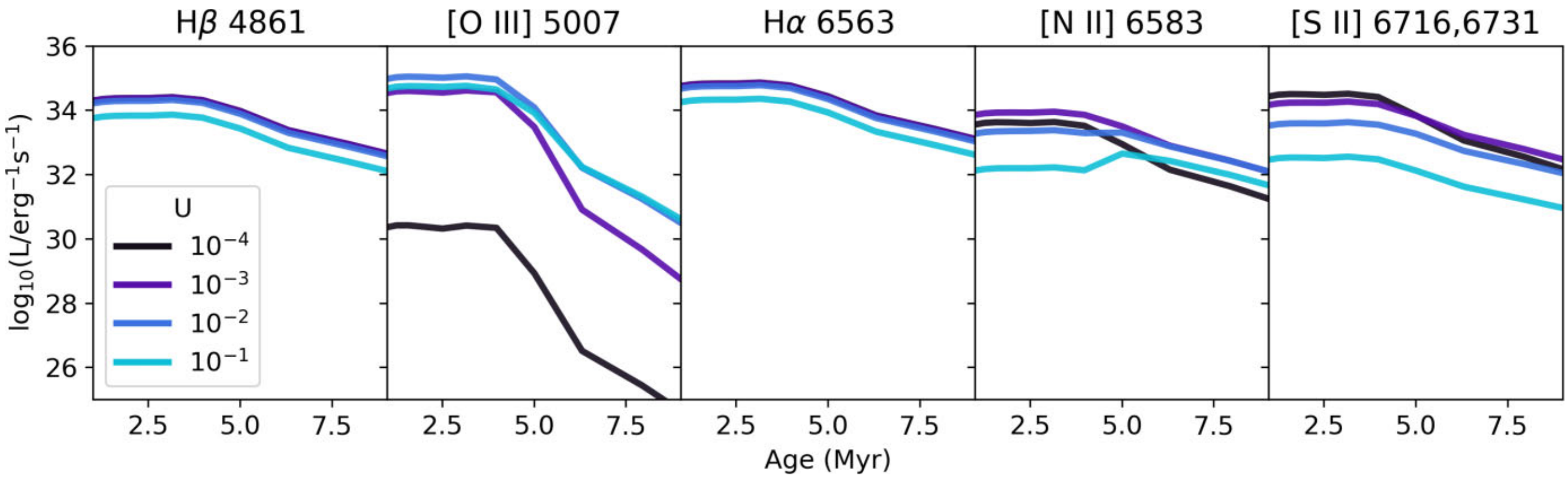

**Figure 9.** Emission strength for the lines H $\beta\lambda$4861, [O III]λ5007, H $\alpha\lambda$6563, [N II]λ6583, and [S II]λ6716,6731 as a function of the age of the model for varying ionization parameter with a fixed metallicity of $Z = 0.014$. Here, the ionization parameter $U$ varies from $10^{-4}$ to $10^{-1}$.

Each of the lines generally show a decrease in luminosity with age, in agreement with the decrease in specific ionizing photon luminosity of the incident radiation with age seen in Fig. 2. For the H $\alpha$ and H $\beta$ lines, we see that increasing the metallicity between $Z = 0.002$–0.02 decreases the luminosity of the line, as expected since lower metallicity models result in hotter H II regions that generate stronger recombination emission (see e.g. M. L. McCall, P. M. Rybski & G. A. Shields 1985; M. A. Dopita & I. N. Evans 1986; S. S. McGaugh 1991; L. J. Kewley & M. A. Dopita 2002; M. A. Dopita et al. 2013). However, H $\alpha$ could increase again as a result of hot stellar phases in old populations (R. Yan & M. R. Blanton 2012; F. Belfiore et al. 2016). The $Z = 0.0003$ model does not follow this trend between metallicity and line strength due to the metallicity being a factor of 10 lower than the other models.

We also observe that increasing the ionization parameter decreases the luminosities of the H $\alpha$ and H $\beta$ lines. The dependence of H $\alpha$ and H $\beta$ on $U$ is contrary to what is seen in S. M. Wilkins et al. 2017 and N. Byler et al. 2017 since here we include dust grains, as outlined in Section 3.1, which affect the thermodynamic properties of the modelled cloud. When dust grains are turned off in CLOUDY, this dependence no longer exists. The dependence on $U$ is small due to the way we normalize our results, as explained in Section 3.1.

For the [O III] line, we observe a much steeper evolution in the line luminosity with age compared to the other lines. At young ages and for a fixed ionization parameter of $U = 10^{-2}$, the [O III] emission is at its maximum when the metallicity is $Z = 0.002$ ($\sim$ 15 per cent of the solar metallicity), unlike the H $\beta$ and H $\alpha$ lines, and in line with the results of Byler with their FSPS models who observe similar but more pronounced trends. As explained in N. Byler et al. (2017), this occurs because oxygen is not sufficiently abundant at the lowest metallicities to produce as much emission as it does at half of the solar metallicity. In Section 4.4, we find that predictions for the [O III] line luminosities vary significantly between different SPS models, which we will discuss further there.

In contrast again to the H $\beta$ and H $\alpha$ lines, the emission from [N II]λ6583 and [S II]λ6716,6731 is weakest for the low metallicity models. We also see that [S II]λ6716,6731 is more dependent on ionization parameter than H $\beta$, H $\alpha$, and [N II]λ6583 is strongest when the ionization parameter is at its minimum, in agreement with other studies such as A. I. Diaz et al. (1991) and L. J. Kewley, D. C. Nicholls & R. S. Sutherland (2019).

In Fig. 10, we show a similar plot to those in Figs 8 and 9 but now with varying hydrogen density. We see that varying the hydrogen density from $n_{\rm H} = 10$ to $n_{\rm H} = 10^5$ cm$^{-3}$ has little impact on the H$\alpha$, H$\beta$ and [S II]$\lambda$6716,6731 line luminosities, in agreement with the results of S. M. Wilkins et al. (2020). We see however that varying the hydrogen density does have an impact on the [O III] line and on the [N II]$\lambda$6583 line at young ages, although the variation in the luminosities is much smaller for [N II]$\lambda$6583. For [O III], the variation is small at young ages and then increases significantly with age.

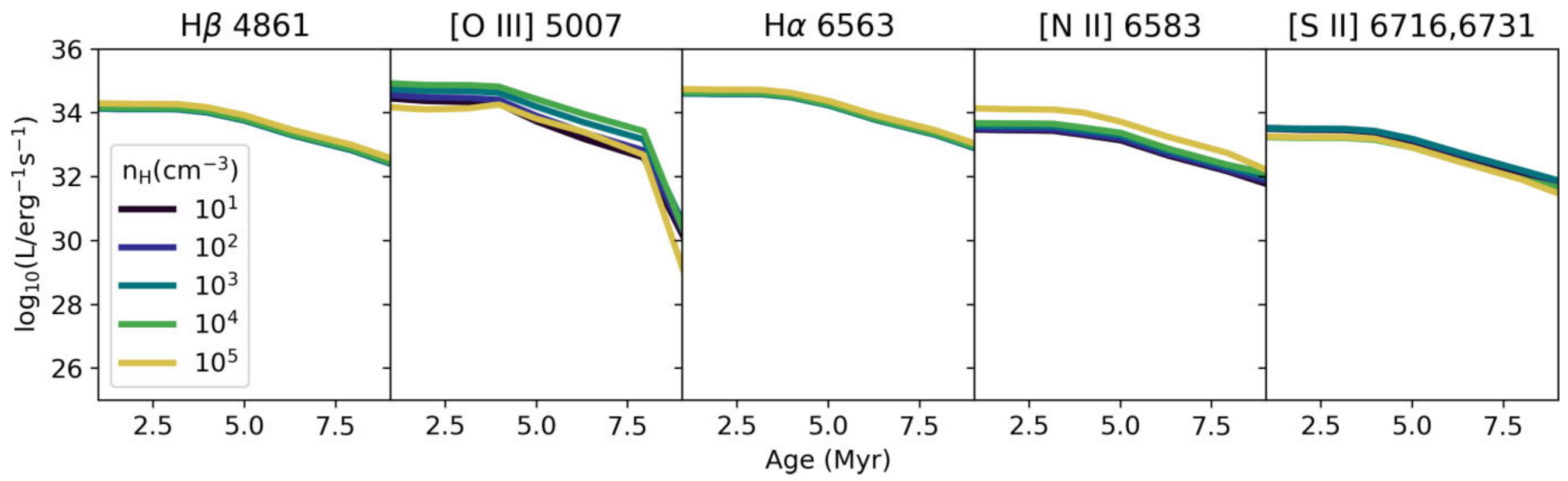

**Figure 10.** Emission strength for the lines H $\beta\lambda$4861, [O III]$\lambda$5007, H $\alpha\lambda$6563, [N II]$\lambda$6583, and [S II]$\lambda$6716,6731 as a function of the age of the model with varying hydrogen density. The hydrogen density $n_{\rm H}$ is varied from $10^1$ to $10^5$ cm$^{-3}$.

## 4.3 Diagnostic diagrams

*JWST* is capable of directly observing auroral emission lines originating from high quantum levels, which are highly sensitive to temperature variations. This sensitivity allows for direct metallicity measurements, as demonstrated by observations of the [O III]$\lambda$4363 line (e.g. R. L. Sanders et al. 2024, I. H. Laseter et al. 2024), along with [S II]$\lambda$4069 and [N II]$\lambda$5755. However, since these require deep observations with long integration times, diagnostic diagrams using ratios of strong emission lines remain useful for obtaining metallicity estimates of large samples of galaxies (R. Maiolino & F. Mannucci 2019). Here, we present the M13 models on these diagrams, and explore the impact of varying ionization parameter, metallicity, and hydrogen density.

Presented in Fig. 11 are the R2, O32, and the R23 ratio plots and their specific definitions and significance will be discussed in the following subsections. We show the M13 models for a fixed solar metallicity population, instantaneous age of 1 Myr, and varying ionization parameter. In these diagrams we plot against the nebular metallicities in the form $\log_{10}(\mathrm{O/H}) + 12$, where we assume a solar value of $\log_{10}(\mathrm{O/H})_{\odot} + 12 = 8.69$ (M. Asplund et al. 2009). On the O32, R23, and N2 diagrams we also show some of the recent *JWST* results, namely measurements by J. R. Trump et al. (2023) of $z \sim 5$–8 galaxies observed in the SMACS 0723 Early Release Observations, measurements by R. L. Sanders et al. (2024) of 16 galaxies at z $= z \sim 2$–9 from the Cosmic Evolution Early Release Science (CEERS) survey (S. L. Finkelstein et al. 2022, 2023), and measurements by I. H. Laseter et al. (2024) at redshifts $z \sim 2$–9 using results from the JADES survey (A. J. Bunker et al. 2023; D. J. Eisenstein et al. 2023; F. D'Eugenio et al. 2024). For [N II]$\lambda$6583 measurements we use those by J. E. Birkin et al. (2023) at $z \sim 4$, B. Welch et al. (2024a, b) at $z \sim 1$–2 and N. S. J. Rogers et al. (2024) at $z \sim 3$. To calculate the nebular metallicities these studies use the 'direct T$_{\rm e}$ method' that uses the ratio of the [O III]$\lambda$4363 and the [O III]$\lambda$4960,5008 doublet to measure the electron temperature of the ISM, then estimates the metallicity from the electron temperature using empirical correlations. These *JWST* candidates all have inferred metallicities that are subsolar ($< 8.69$). There is currently a lack of data in the high metallicity regime at high redshift to compare our models to.

### 4.3.1 O32

First, we show the O32 ratio in Fig. 11, which is defined as the ratio log([O III]$\lambda$5007/[O II]$\lambda$3726,3729)). The O32 ratio is primarily used to measure the ionization parameter since it is a ratio of two ionization states of the same element (e.g. A. I. Díaz et al. 2000; L. J. Kewley & M. A. Dopita 2002; T. Nagao, R. Maiolino & A. Marconi 2006), and we see this clearly in Fig. 11 in the clear separation between the lines for different ionization parameters and how the $z \sim 2$–9 *JWST* observations of young galaxies favor the models with high ionization parameters around $U \approx 10^{-2}$. In the work by A. I. Díaz et al. (2000), they use CLOUDY to obtain a grid of photoionization models and derive the following expression for the O32 diagnostic as

$$\log U = -0.80 \log([\mathrm{O_{II}}]/[\mathrm{O_{III}}]) - 3.02 \quad . \tag{12}$$

for metallicities $12 + \log(\mathrm{O/H}) = 7.2$–8.8. We however, for a metallicity of $12 + \log(\mathrm{O/H}) = 7$, find the quadratic relationship

$$\log([\mathrm{O_{III}}]/[\mathrm{O_{II}}]) = -3160U^2 + 357U - 2.10 \quad . \tag{13}$$

In Fig. 11, there is not a clear relationship between the hydrogen density $n_{\rm H}$ and the O32 ratio, and some of the *JWST* points lie above the regime of the $n_{\rm H} = 10^1$–$10^5$ cm$^{-3}$ models.

### 4.3.2 N2

The N2 ratio is defined as [N II]$\lambda$6583/H$\alpha$. In Fig. 11, we can see a strong dependence on the ionization parameter and the metallicity. As seen earlier in Figs 8 and 9, there is generally more [N II]$\lambda$6583 emission for higher metallicities and lower ionization parameters. The relation between $n_{\rm H}$ and the N2 ratio in Fig. 11 is more clear than that of the O32 ratio, with the higher hydrogen density models leading to high N2 ratios. Many of the *JWST* points lie close to the models, particularly the model with the highest hydrogen density of $n_{\rm H} = 10^5$ cm$^{-3}$. Unlike the O32 ratio, which primarily traces the ionization parameter, the N2 ratio is more sensitive to nitrogen abundance and due to its quadratic dependence on metallicity, the N2 ratio can drop significantly in low-metallicity galaxies such as those observed by

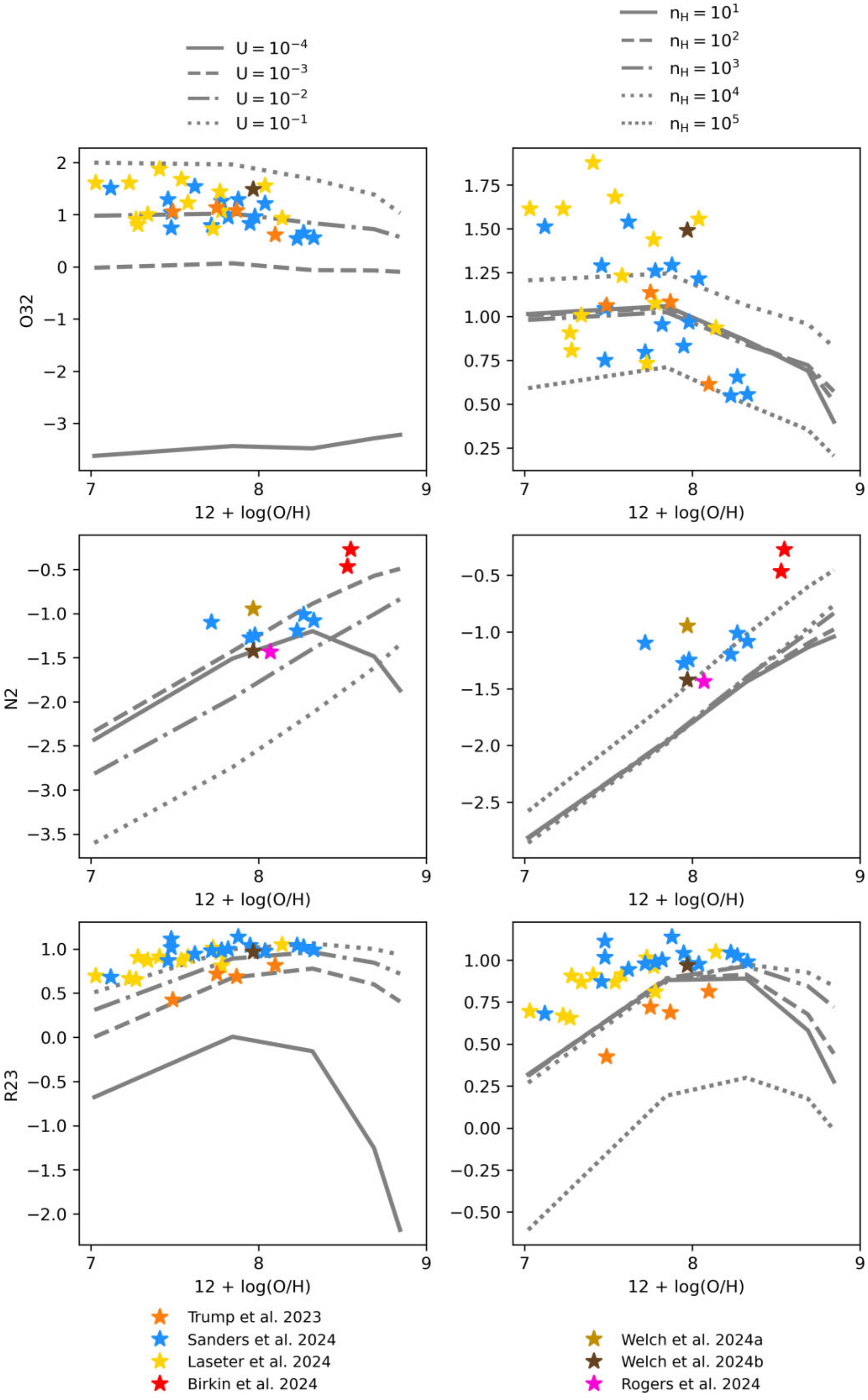

**Figure 11.** Our results for the O32, N2, and R23 ratios for different ionization parameters $U$ and hydrogen densities $n_{\rm H}$ compared with *JWST* observations. These ratios are defined as log([O III]λ5007/[O II]λ3726,3729), [N II]λ6583/H α, and log(([O II]λ3726,3729 + [O III]λ4959,5007)/H β)), respectively. The results for varying ionization parameters and hydrogen densities are shown on the left and right respectively. The *JWST* observations are marked with stars, as indicated in the legend.

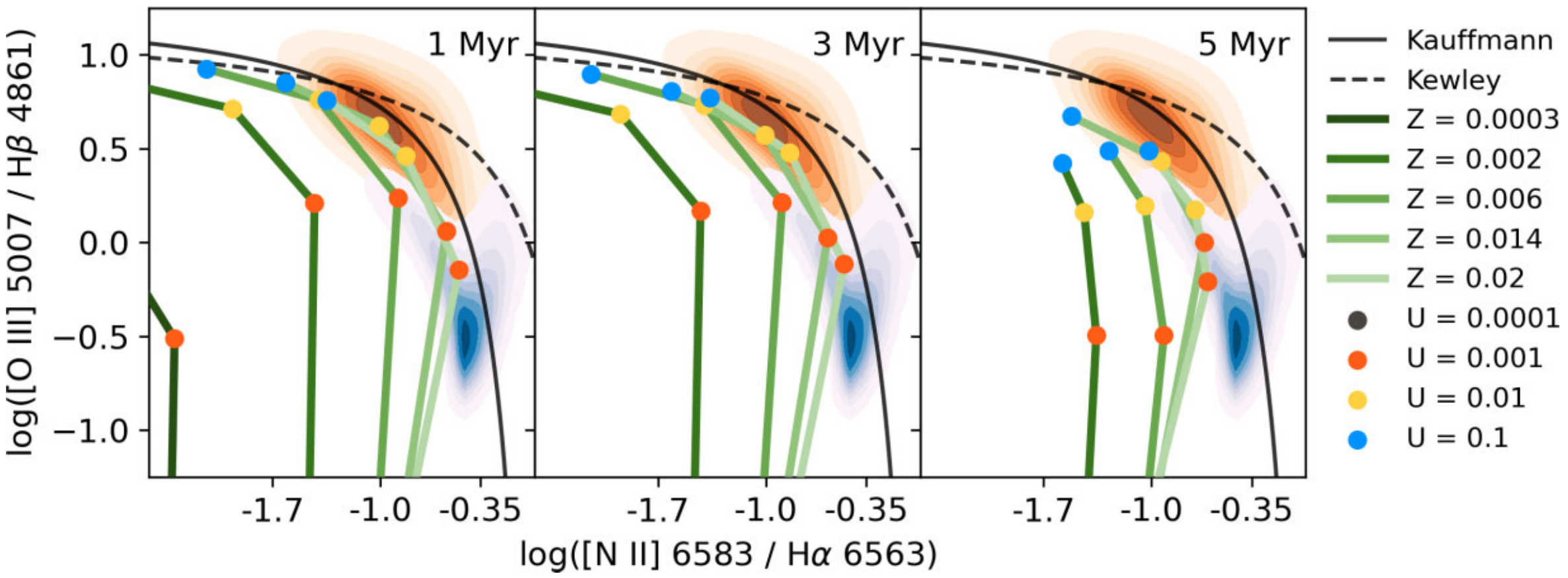

**Figure 12.** Our results for ages 1, 3, and 5 Myr on the Baldwin–Phillips–Terlevich (BPT) diagram, a plot of the ratio [O III]λ5007/H β λ4861 against the ratio N IIλ6583/H αλ6563. Each panel corresponds to a different age. Lines represent fixed metallicities, while points indicate varying ionization parameters. Additionally, the G. Kauffmann et al. (2003) and L. J. Kewley et al. (2006) lines are shown, along with data from SDSS DR8 (shown as the lower contour) and JADES DR3 (shown with the higher contour).

*JWST* – reaching values as low as $10^{-2}$ – making the [N II]λ6583 line extremely faint in such environments (R. Maiolino & F. Mannucci 2019).

### 4.3.3 R23

R23 is defined as the ratio of the main ionization stages of oxygen (= log(([O II]λ3726,3729 + [O III]λ4959,5007)/H β)), making it less sensitive to the ionization structure of H II regions. As seen in Fig. 11, R23 depends on the ionization parameter and exhibits a double-branched structure, meaning a single R23 value can correspond to two distinct metallicities (L. S. Pilyugin & T. X. Thuan 2005; L. J. Kewley & S. L. Ellison 2008). The *JWST* observations on the R23 diagram don't align with a particular model as clearly as on the O32 diagram, but it is still clear that the higher ionization parameter models are preferred to the lower $U = 10^{-4}$ model. Like the O32 ratio, there is not a linear relation between hydrogen density and the R23 ratio in Fig. 11.

Very high R23 ratios have been observed in some *JWST* galaxies, as seen in Fig. 11, and we infer that *JWST* galaxies consist of very young stellar populations, and exhibit high ionization parameters.

### 4.3.4 BPT

The Baldwin–Phillips–Terlevich (BPT) diagram, a plot of the ratio [O III]λ5007/H β λ4861 against the ratio N IIλ6583/H αλ6563 is commonly used to distinguish star-forming H II regions from areas excited by other mechanisms, such as AGN. To separate these regions, the lines by G. Kauffmann et al. (2003) and L. J. Kewley et al. (2006) are used. Both of these separation boundaries are derived from empirical SDSS libraries.

We choose to show our results on the BPT diagram alongside data from local star-forming galaxies in the Eighth SDSS Data Release (DR8; D. G. York et al. 2000; H. Aihara et al. 2011) as well as from the third data release of JADES, the *JWST* Advanced Deep Extragalactic Survey (A. J. Bunker et al. 2023; D. J. Eisenstein et al. 2023; F. D'Eugenio et al. 2024). The SDSS data has a maximum redshift of $z = 0.4$ and has a total of 246 980 galaxies that have detections of the lines [O III]λ5007, Hβ, [N II]λ6583, and H α. JADES reaches a maximum redshift of $z = 12.5$ however after applying a filter to ensure the redshift is determined from at least one emission line using the medium-resolution grating, and to improve data quality, only 150 values remain for plotting on the BPT diagram due to a lack of [N II] measurements. In contrast, there are 1849 measurements for [O III].

We show our results for different metallicities and ionization parameters in Fig. 12 alongside the data and the Kauffmann and Kewley lines. The JADES results are shown as orange contours and the SDSS results are shown as blue contours. Results for different ages are shown in separate panels where lines represent fixed metallicities and points indicate different ionization values. Not visible on the diagram are the results of the $U = 10^{-4}$ model since it resulted in much smaller [O III]/H β values than the other values shown.

We see that the lowest metallicity models have smaller nitrogen abundances due to secondary nitrogen production and therefore lower [N II] fluxes as expected. Increased population ages lead to lower [O III]/H β values and higher [N II]/H α values. Comparing our results to the data contours, we see that the JADES and SDSS galaxies are best represented by our model for a 1–3 Myr population with a metallicity of around $Z = 0.01$–$0.02$ and a high ionization parameter between $U = 10^{-2}$–$10^{-1}$. The local star forming galaxies on the other hand are better represented by a 1–3 Myr population model with the same metallicity of $Z = 0.01$–$0.02$ but with a lower ionization parameter of $U = 10^{-3}$–$10^{-4}$. The super-solar metallicity model aligns most closely with the data for each of the three ages, likely due to the inclusion of rotation in the model. This outcome contrasts with the M13 models, where the $Z = 0.02$ model has much lower line strengths at 5 Myr and the lower metallicity $Z = 0.01$ model provides a better fit. For further details, refer to Appendix C.

Different combinations of age, metallicity, and ionization parameter can lead to the same point in the BPT diagram, and it is important to remember that the galaxies studied by SDSS and JADES cannot be ideally categorized by a single stellar population as composite populations are often present. Although at these ages, composite populations and SSPs should be very similar. Additionally, some JADES galaxies lie above the Kewley and Kauffmann lines, which may suggest the presence of AGN contributing to the line luminosities (D. D. Kocevski et al. 2023; M. Habouzit 2024).

Studies of line emission in galaxies at redshift > 3 (D. Masters et al. 2014; C. C. Steidel et al. 2014; A. E. Shapley et al. 2015; A. L. Strom et al. 2017; A. L. Faisst et al. 2018) have found that increasing redshift increases the [O III]/H $\beta$ ratio for a given value of [N II]/H $\alpha$ on the BPT diagram. However, R. L. Sanders et al. (2023) analysed *JWST*/NIRSpec observations of 164 galaxies at $z = 2.0$–9.3 from the CEERS survey and showed that galaxies at $z = 2.7$–6.5 fall approximately in the same region on the BPT diagram as $z =$ 2.0–2.7 galaxies, suggesting that ISM ionization conditions do not significantly change between z $\sim$ 2 and z $\sim$ 6. Our results do not assume a redshift but the first set of results from literature imply that the data contours in Fig. 12 would shift upwards with increasing redshift and align with even younger age models, especially for solar metallicity.

Many other studies have also studied how ionization conditions can impact the positions of galaxies on these diagnostic diagrams. For example, P. Garg et al. (2022) found that reducing the ionization parameter in their nebular line emission models shifts galaxies toward lower [O III] and higher [N II] positions on the BPT diagram, which aligns with our findings presented in Fig. 12; as a result, they find that coupling galaxies from the cosmological hydrodynamical simulation SIMBA (R. Davé et al. 2019) with lower ionization parameter models results in better agreement with the SDSS-DR8 data. Similarly, P. Garg et al. (2022) find that decreasing the hydrogen density also moves the galaxies toward lower [O III] and higher [N II], but by a smaller extent than when decreasing the ionization parameter. They also found that increasing the slope of the ionizing spectra results in a shift toward higher [O III] and lower [N II] values, thus moving the distribution away from the SDSS observations.

We have shown our results for simple stellar populations on various diagnostic diagrams (N2, O32, R23, BPT) in Fig. 11, alongside recent *JWST* measurements and SDSS data for these strong line ratios. These diagrams provide insights into how the ionization conditions and other physical properties that we have modelled compare to those of high redshift galaxies detected by *JWST*. In particular we see that our youngest models, with high ionization parameters and metallicities around $Z \approx 0.02$, lie closest to the observations. Increasing the age would reduce the [O III] line strengths (see Fig. 8) significantly and cause the $y$-axis ratio of the BPT diagram in Fig. 11 to decrease due to the sensitivity of the [O III] line. This would shift the theoretical lines in Fig. 12 away from the *JWST* data, causing the lines to move downward, further diverging from the SDSS data as well.

## 4.4 Comparison to other models

In this section, we compare our predictions for the nebular line emission to that obtained from the widely used models of BC03 (G. Bruzual & S. Charlot 2003), BPASS (version 2.2.1[6], J. J. Eldridge et al. 2017; E. R. Stanway & J. J. Eldridge 2018), FSPS (version 3.2 C. Conroy et al. 2009c; C. Conroy & J. E. Gunn 2010), and Starburst99 (SB99, C. Leitherer et al. 1999) as an ionization source. For consistency, we run all of the SPS models through CLOUDY with the same modelling parameters as discussed in Section 3.1. Before comparing the models, we first briefly summarize their main properties such as their chosen stellar tracks below:

*BC03:*
BC03 combines the library of low- and intermediate-mass Padova tracks by L. Girardi et al. (2000) with high-mass tracks from the older 'Padova 1994' library to create an updated library, referred to as the 'Padova 2000' library, which covers a full range of initial stellar masses. They then also combine this with the 'Geneva' library, a set of tracks for solar metallicity (compiled by G. Schaller et al. 1992a; C. Charbonnel et al. 1996, 1999). The IMF for this model is Chabrier with a maximum mass of 100 $M_{\odot}$.

*BPASS:*
Unlike the other models, BPASS accounts for mass-loss or gain due to binary interactions and also tracks the evolution of binary separation and orbital angular momentum. BPASS is a descendant of the STARS code (L. G. Henyey, J. E. Forbes & N. L. Gould 1964; P. P. Eggleton 1971; J. J. Eldridge, R. G. Izzard & C. A. Tout 2008) and uses isocontours rather than isochrones since there are two stars interacting. We run both the BPASS models that include binary interactions and those that do not to observe the impact of binarity clearly. We choose to use the BPASS model with broken power law slopes of $\alpha_1 = -1.30$ and $\alpha_2 = -2.35$ for the low- and high-mass regimes of the initial mass function respectively, and a maximum mass of 100 $M_{\odot}$.

*FSPS:*
FSPS gives the option to choose which set of stellar tracks are used; we chose the MIST isochrones (J. Choi et al. 2016; A. Dotter 2016), which include stellar rotation like our M24 models. The chosen FSPS model has a Chabrier IMF that is defined over $0.08 \leq M/M_{\odot} \leq 120$.

*SB99:*
The focus of SB99 has historically been on relatively massive stars and young starbursts and was computed using the evolutionary tracks of the Geneva group for high mass-loss rates. We use the original 1999 instantaneous burst star formation models without the nebular continuum added for self-consistency, assuming a power-law initial mass function (IMF) with an exponent of $\alpha = -2.35$ for the mass range $1 \leq M/M_{\odot} \leq 100$, approximating the classical Salpeter IMF.

All of the models use isochrones that take into account Wolf–Rayet (WR) stars in their calculations in some way. The isochrones used by BPASS and FSPS identify a star as a WR and use the WR mass-loss prescription by T. Nugis & H. J. G. L. M. Lamers (2000) if the star has a surface temperature above $10^4$ K and has a surface hydrogen mass fraction below 40 per cent. The isochrones from G. Schaller et al. (1992b), which are part of the Geneva set of isochrones used by BC03, M13, and SB99, identify WR stars in the same way but apply a modified wind prescription based on the models of J. I. Castor, D. C. Abbott & R. I. Klein (1975) and R. P. Kudritzki et al. (1989). SB99 favors the enhanced mass-loss versions of the Geneva isochrones to provide a more accurate representation of WR star properties.

Most of the SPS models, including M13 and M24, were constructed by drawing synthetic spectra from the BaSeL stellar library (T. Lejeune et al. 1997), which was obtained by merging the theoretical Kurucz library (R. L. Kurucz 1979 and revisions) of model atmospheres with atmospheres for cooler stars. Some of these models choose to supplement the BaSeL library with other libraries. For example, BPASS choose to use the empirical, high resolution stellar atmosphere models provided by C. K. Conroy that are known as the C3K models (see C. Conroy, G. J. Graves & P. G. van Dokkum 2014). BPASS also supplement the C3K library with the theoretical Potsdam Wolf–Rayet (PoWR, A. Sander et al. 2015) atmosphere models for WR stars. BC03 also use the non-LTE atmospheric models from T. Rauch (2002) to describe the radiation of WR stars for solar ($Z = Z_{\odot}$) and 10 per cent solar metallicity ($Z = 0.1Z_{\odot}$). These models account for metal-line absorption from elements ranging from hydrogen to nickel. SB99 supplement the BaSeL library with theoretical atmospheres from W. Schmutz, C. Leitherer & R. Gruenwald (1992) for WR stars. In Table 2, we summarize the stellar input tracks and stellar atmosphere models that different SPS models use (for ages < 10 Myr).

[6]We choose not to use the newer version 2.3 to ensure consistency between the initial mass functions used by the models being compared.

**Table 2.** Stellar input tracks and spectral libraries used by the M13, M24, FSPS, BPASS, BC03, and SB99 SPS models for ages less than 10 Myr, as shown in Figs 13 and 14. Only libraries extending to the UV that can impact the ionizing photon production rate are listed.

| SPS model | M13 | M24 | FSPS | BPASS | BC03 | SB99 |
|---|---|---|---|---|---|---|
| Stellar input tracks (isochrones) | Geneva | Updated Geneva (with rotation) | MIST | Isocontours from a descendant of the STARS code | Padova + Geneva | High mass-loss Geneva |
| Spectral libraries | BaSeL | BaSeL | BaSeL | C3K, supplemented with WR | STELIB, BaSeL, Pickles + non-LTE | BaSeL, supplemented with theoretical WR |

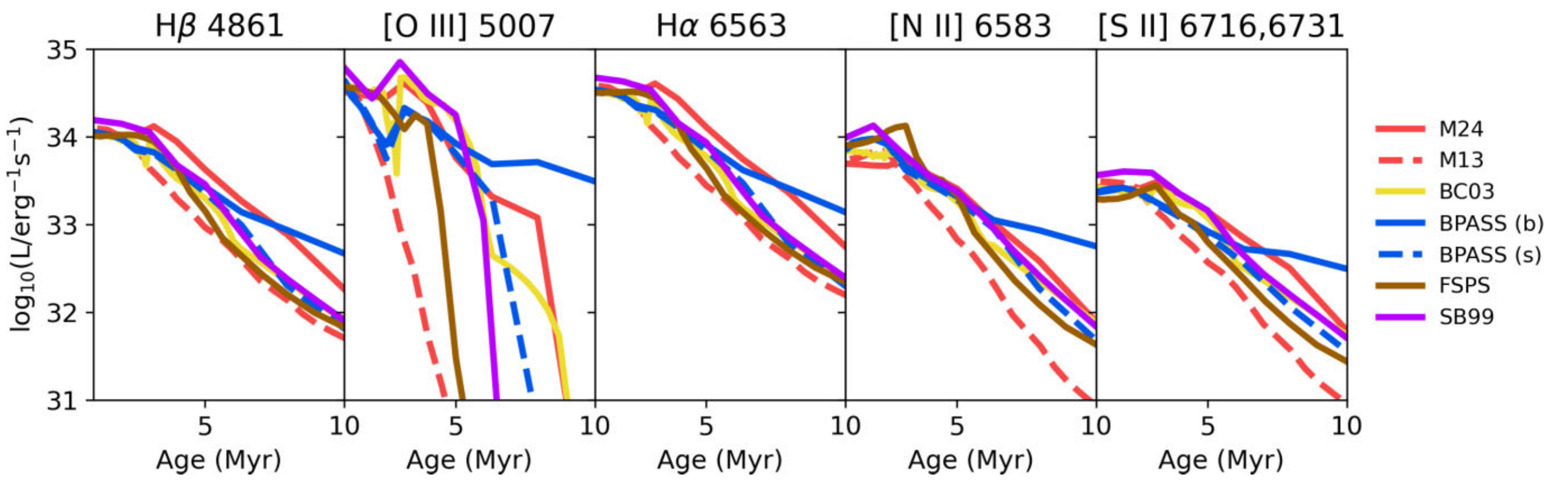

**Figure 13.** Emission strength for the lines H $\beta\lambda$4861, [O III]$\lambda$5007, H $\alpha\lambda$6563, [N II]$\lambda$6583, and [S II]$\lambda$6716,6731 for different SPS models at solar metallicity ($Z = 0.014$): M13, M24, BC03, BPASS including binary (b) interactions, BPASS with single (s) stars and no binary interactions, FSPS, and Starburst99 (SB99).

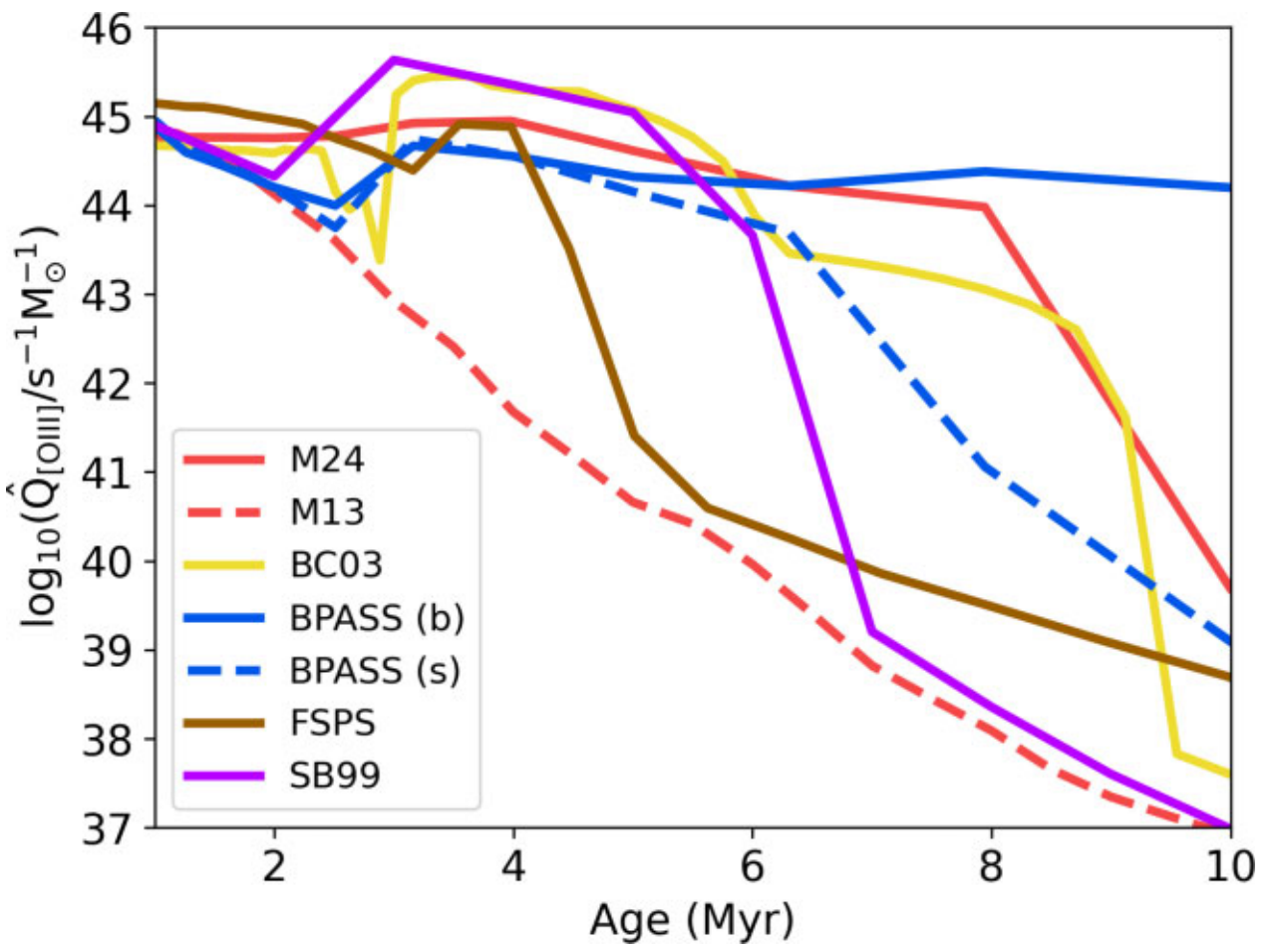

**Figure 14.** The ionizing photon production rate $\hat{Q}_{[\mathrm{O\,III}]}$ for different SPS models at solar metallicity ($Z = 0.014$): M13, M24, BC03, BPASS including binary (b) interactions, BPASS with single (s) stars and no binary interactions, FSPS, and Starburst99 (SB99).

In works such as C. Maraston & G. Strömbäck (2011) and P. R. T. Coelho, G. Bruzual & S. Charlot (2020), the differences between theoretical and empirical libraries are evaluated and the impact of limited versus full HR coverage is compared. Whether a library is theoretical like BaSeL, empirical like C3K or a combination may well be another factor that leads to differences in ionizing photon production rates. C. M. Byrne & E. R. Stanway (2023) also find that the choice of spectral library impacts their results when they vary the library with a fixed BPASS model, and show that line indices can vary strongly between different templates.

We compare line luminosities predicted for each SPS model from 1 to 10 Myr in Fig. 13 for the lines H$\beta$, H$\alpha$, [O III]$\lambda$5007, [N II]$\lambda$6583, and [S II]$\lambda$6716,6731. We find that the majority of lines show little variation in luminosities across different models. However, the [O III] line luminosities exhibit significant differences. For models that do not account for binary effects, [O III] luminosities decrease dramatically at specific ages unique to each model. The [O III] luminosities drop below $10^{31}$ erg s$^{-1}$ first in the M13 model at 5 Myr, followed by the FSPS model at 5.6 Myr, SB99 at 7 Myr, the single-star BPASS model at 7.9 Myr, the BC03 model at 9.1 Myr, M24 at 10 Myr, and finally the binary BPASS at 126 Myr. Interestingly, our newly explored M24 offers relatively high [O III] just for single stars and before the addition of binary effects.

To explore the reasons behind the varying [O III] line luminosities, we present Fig. 14, which plots the [O III] ionizing photon production rate against age for each model. As seen, the M24 model starts at a later age than the M13 model because the rotating isochrones have a different set of ages compared to the original isochrones used in M13. We find that the production rate of photons capable of ionizing $O^+$ to $O^{2+}$ varies significantly between SPS models, resulting in the different [O III] line luminosities observed in Fig. 13.

S. M. Wilkins et al. (2023) show that the median [O III] equivalent widths predicted by the FLARES zoom-in simulations (C. C. Lovell et al. 2021; A. P. Vijayan et al. 2021), when coupled to the BPASS SPS model, are consistent with spectroscopic constraints from *JWST* (F. Sun et al. 2022; J. Matthee et al. 2023). However, the simulation results do not predict the tail of galaxies exhibiting extremely high equivalent widths greater than 2000Å, as identified by these *JWST* studies. Potential explanations for this discrepancy may include limitations in the available observational sample, the star formation efficiency used in the simulation, or an important, unaccounted source of ionizing photons in high redshift galaxies that is not captured by the current SPS or AGN models.

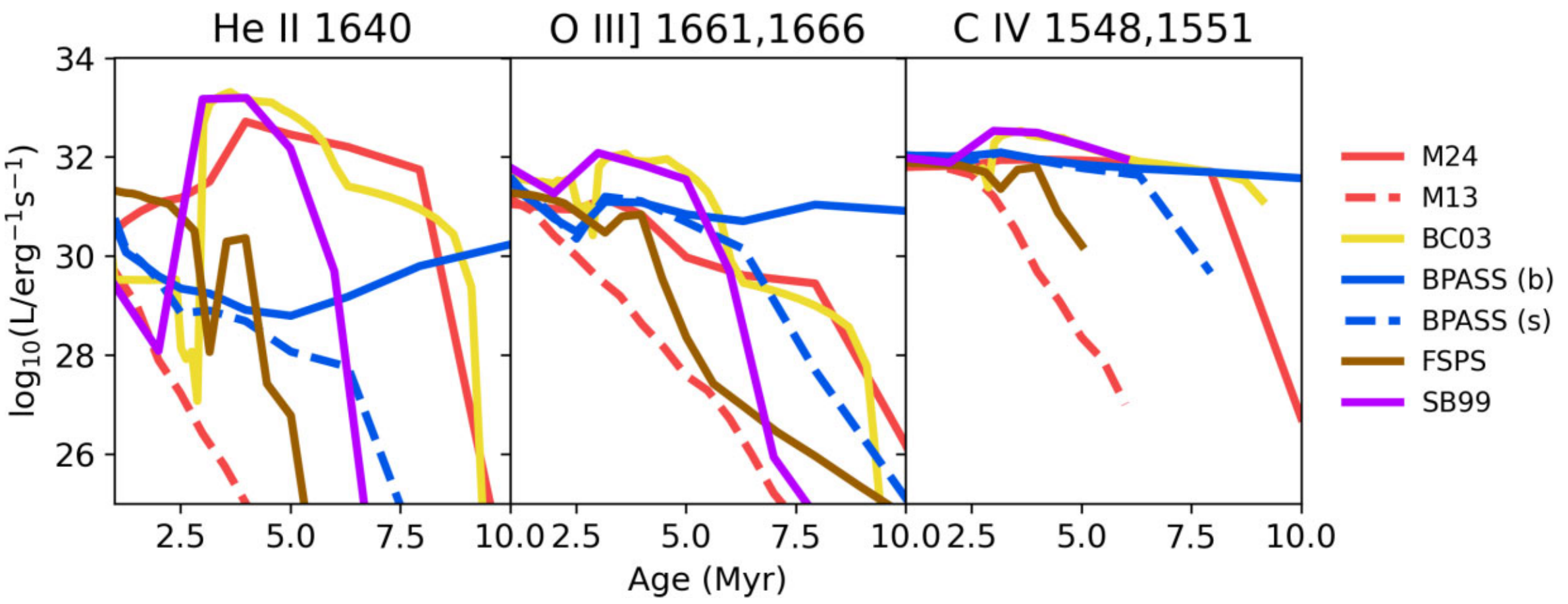

**Figure 15.** Emission strength for the lines He II$\lambda$1640, O III]$\lambda$1661, 1666 and C IV$\lambda$1548, 1551 for different SPS models at solar metallicity ($Z = 0.014$): M13, M24, BC03, BPASS including binary (b) interactions, BPASS with single (s) stars and no binary interactions, FSPS, and Starburst99 (SB99). For C IV$\lambda$1548, 1551, the plotted lines are truncated at the ages where equivalent widths drop to zero.

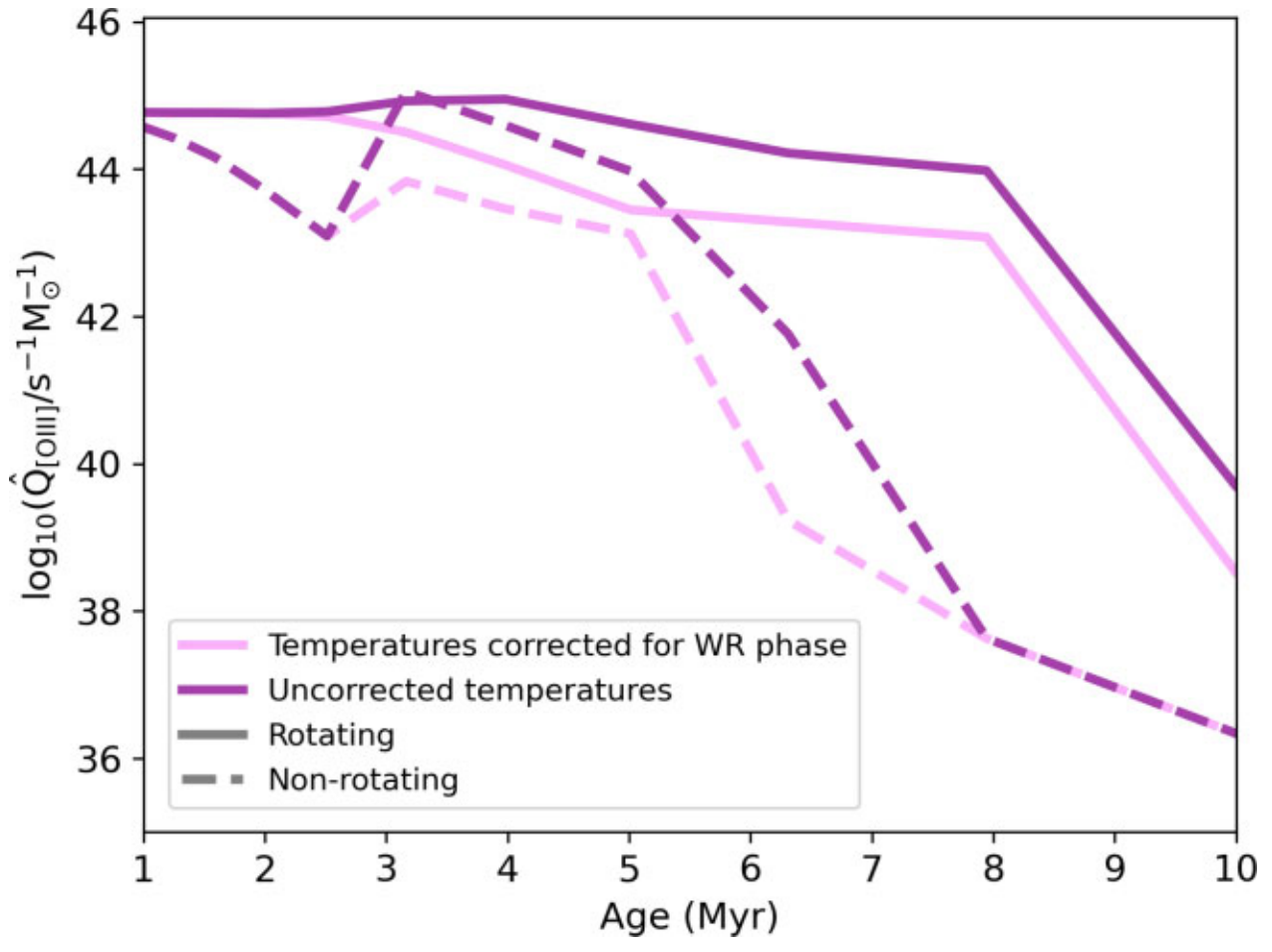

**Figure 16.** The specific ionizing photon luminosity for [O III] of the M24 models at solar metallicity ($Z = 0.014$) as a function of age for different rotations and methods of calculating the temperature: corrected or uncorrected for WR winds.

Given that the [O III]$\lambda$5007 line corresponds to a high ionization energy, we also compare the line luminosities of UV lines with high ionization energies $E$ such as He II$\lambda$1640 ($E \approx 54$ eV), O III]$\lambda$1661, 1666 ($E \approx 35$ eV), and C IV$\lambda$1548, 1551 ($E \approx 48$ eV) across the different SPS models, as shown in Fig. 15. As anticipated, the luminosities of these lines also exhibit significant variation between the models, primarily due to the greater uncertainties in the wavelength range necessary for calculating ionizing photon production rates. As seen here, the M13 model does not result in significant He II$\lambda$1640 emission. However, while the M13 model generates significantly less He II than the other models, it produces more comparable levels of O III]$\lambda$1661, 1666 and C IV$\lambda$1548, 1551 emission. For He II$\lambda$1640, the M24 model again results in higher line luminosities than M13 and even higher compared to the BPASS model with binary interactions, and has line strengths comparable to the other models for the O III]$\lambda$1661,1666 and C IV$\lambda$1548,1551 lines.

To further demonstrate the effect of rotation as shown by the M24 model, we show Figs 16 and 17. At 8 Myr in Fig. 16, and for uncorrected effective temperatures, the model with rotation produces $\approx 10^{44}$ more [O III] ionizing photons per second, resulting in higher [O III] luminosities in Fig. 17. We also observe the impact of whether the effective temperatures are corrected for the winds of WR stars: at 5 Myr and without rotation, the uncorrected model has a [O III] ionizing photon production rate that is $\approx 7$ times greater than the model that has been corrected. Before the WR phase begins around 3 Myr, the uncorrected and corrected models are identical.

We have found significant differences between models with different input stellar physics, particularly in the line luminosities of high ionization lines due to vast differences in the ionizing photon production rate as seen in Fig. 14 for [O III] where the models differ by a maximum of $\approx 10^{44}$ s$^{-1}$ M$_\odot^{-1}$. These discrepancies highlight the need for careful selection of models when interpreting observations with strong UV emission, such as from He II$\lambda$1640 produced by stellar populations with low metallicities (e.g. A. E. Shapley et al. 2003; P. Cassata et al. 2013; D. A. Berg et al. 2018; A. Saxena et al. 2020; X. Wang et al. 2024), and observations from *JWST* with strong [O III] emission (e.g. F. Sun et al. 2022; H. Katz et al. 2023b; G. Roberts-Borsani et al. 2024; A. Saxena et al. 2024).

Rotating stars generally exhibit harder ionizing spectra and greater luminosities, which contributes to the M24 model with rotation producing one of the highest [O III] line luminosities. FSPS also includes rotation, but results in a slightly lower [O III], indicating the effect of varying the rotation parameter in stellar tracks. The inclusion of binary effects in the BPASS model will also impact the hardness of the spectra: for example P. Liu et al. (2024) show that the BPASS binary model produces $\sim 40$ % more ionizing photons than the BPASS model with just single stars, and we too see a large difference between these models in Fig. 14. It is possible then that a model that incorporates both stellar rotation and binary interactions would lead to even higher ionizing photon production rates than models that consider only rotation or binarity.

The WR phase, that all the models include in their calculations in some way, is short, typically occurring around $\sim 3$ Myr and lasting only a few Myr. Indeed, near this age we observe an increase in the number of [O III] ionizing photons being produced for M24, BC03, FSPS, and SB99, with the FSPS peak likely being delayed since rotation increases the WR star lifetime. We don't see an obvious WR contribution from the binary BPASS model, but it is observed much more clearly in the BPASS single model. Unlike FSPS and BPASS, the increase in flux due to WR stars in the BC03 model lasts longer than a few Myr. We do not see obvious contributions from these stars in the M13 model because the Geneva tracks used lead to a considerably lower number of WR stars than the Padova tracks (as observed by S. Charlot 1996 and G. Bruzual & S. Charlot 2003). This demonstrates how factors such as the inclusion of WR stars in stellar population modeling significantly influences the ionizing spectrum.

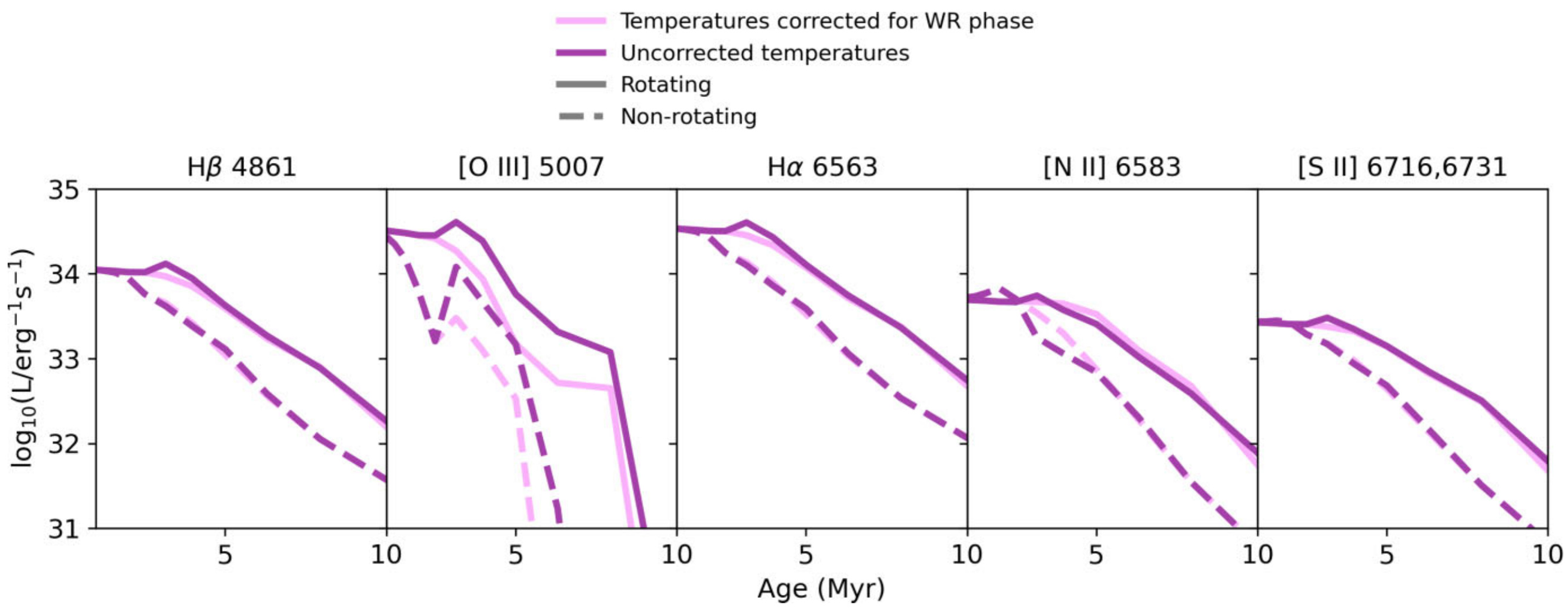

**Figure 17.** Emission strengths for the lines H $\beta\lambda$4861, [O III]$\lambda$5007, H $\alpha\lambda$6563, [N II]$\lambda$6583, and [S II]$\lambda$6716,6731 as a function of age for solar metallicity for different rotations and methods of calculating the temperature: corrected or uncorrected for WR winds.

In this work, we adopted the Maraston model which offers a unique treatment of the TP-AGB phase. For intermediate-age (0.6–2 Gyr) populations in $z \approx 2$ spectra, this phase has a significant impact at near-IR wavelengths observable by *JWST* (C. Maraston et al. 2006; C. Tonini et al. 2010). Specifically, we used the new M24 model which improves upon the original M05 by refining the calibration of the onset age and fuel consumption for the TP-AGB phase as well as using the latest stellar tracks. Recent findings by S. Lu et al. (2024) show that the TP-AGB prescription in the M13 models produces the highest fraction of detected TP-AGB features, such as TiO and CN, in full visible and near-IR rest-frame spectra of quiescent galaxies at redshifts $z \sim 1$–2, as observed by *JWST*. Moreover, the TP-AGB in stellar population models is also needed to reproduce the spectra of star forming galaxies (e.g. R. Riffel et al. 2015) therefore a proper treatment of this stellar phase has implications on a range of star formation histories and redshift. While this study focuses on stellar populations younger than 10 Myr, the impact of the TP-AGB phase becomes more pronounced at older ages, which could lead to different outcomes when comparing models.

Our comparison with widely used SPS models – BC03, BPASS, FSPS, and SB99 – has revealed significant discrepancies, particularly in predicting the luminosities of high-ionization lines. These differences are driven by assumptions about the WR phase and variations in ionizing photon production rates. The choice of SPS models is critical, as different models can lead to varying predictions when coupled with cosmological hydrodynamical simulations of emission line galaxy populations and in SED fitting, where each model assigns a unique spectrum for the same input parameters. Observations will be key in determining which predictions best match observed properties.

## 5 CONCLUSIONS

We present, for the first time, the newly calculated M24 stellar population model, which incorporates the latest Geneva tracks with rotation (S. Ekström et al. 2012; C. Georgy et al. 2013; J. H. Groh et al. 2019; P. Eggenberger et al. 2021; N. Yusof et al. 2022). This model has been integrated with the Cloudy photoionization code to simulate the physical conditions within a gas cloud to obtain the nebular continuum and line emission. We have presented a detailed analysis of the ionization conditions and physical properties of our models, using various diagnostic diagrams (N2, O32, R23, BPT) and comparing these results to recent *JWST* measurements and SDSS data. Our specific conclusions are as follows:

(i) We find that incorporating rotation and different effective temperatures during the Wolf–Rayet phase in the M24 model introduces significant differences compared to the non-rotating M13 model. For instance, at 5 Myr, non-rotating models using uncorrected effective temperatures result in ionizing photon production rates $\hat{Q}_{[\mathrm{OIII}]}$ that are $\approx 10$ times greater compared to those using effective temperatures corrected for WR winds. In models using uncorrected effective temperatures, the rotating model produces [O III]$\lambda$5007 luminosities approximately $\approx 10^6$ times greater than the non-rotating model at 8 Myr.

(ii) We found that our youngest M24 models with high ionization parameters between $U = 10^{-1}$–$10^{-2}$ best represent the observed data from *JWST*. Increasing the age of the population from 3 to 5 Myr significantly alters the [O III]$\lambda$5007 line strengths in the BPT diagram, which causes a deviation from the *JWST* data, reinforcing the idea that younger stellar populations align more closely with current observations of galaxies at high redshift with strong nebular emission. We also find that the M24 models with metallicities $Z = 0.014 - 0.02$, lie closest to the *JWST* data, potentially identifying metal-rich gas at high redshift.

(iii) The comparison of our results with other stellar population synthesis (SPS) models, including BC03, BPASS, and FSPS shows relatively good agreement for the strengths of the lines H $\beta$, H $\alpha$, [N II]$\lambda$6583 and [S II]$\lambda$6716,6731.

(iv) We find significant differences in the production rates of hard ionizing photons, capable of producing [O III]$\lambda$5007, He II$\lambda$1640, O III]$\lambda$1661, 1666, and C IV$\lambda$1548, 1551] emission. The models differ by a maximum of $\Delta\hat{Q}_{[\mathrm{OIII}]\lambda 5007} \approx 10^{44}$ s$^{-1}$ M$_{\odot}^{-1}$ for [O III]$\lambda$5007. These discrepancies arise from variations in stellar input tracks, spectral libraries, and other assumptions used in these models. For instance, BC03, SB99 and M24 show the highest [O III]$\lambda$5007 ionizing photon production rates, likely due to the inclusion of non-LTE atmospheres, enhanced mass-loss isochrones, and rotation, while the M13 model show the lowest. The M24 model, which accounts for rotation, exhibits higher [O III]$\lambda$5007 production rates, aligning more closely with those of other SPS models compared to the M13 model, which does not include rotation. The M24 model also results in equivalent widths as high as models that include binary evolution, highlighting the importance of understanding the model input physics.

## ACKNOWLEDGEMENTS

SN gives special thanks to Daniel Ballard for the outstanding pun in the paper title. We thank Sylvia Ekström, for support in the use of the Geneva tracks and for specific calculations, and George Meynet for our discussions.

Numerical computations were done on the Sciama High Performance Compute (HPC) cluster which is supported by the Institute of Cosmology & Gravitation (ICG), SEPNet and the University of Portsmouth. We acknowledge the use of the NUMPY (C. R. Harris et al. 2020), MATPLOTLIB (J. D. Hunter 2007), and SCIPY (P. Virtanen et al. 2020) packages. This work was supported by the Science and Technology Facilities Council (STFC) through a PhD studentship [2903874].

## DATA AVAILABILITY

All of the models and code used in this work are publicly available at the following places online:

(i) The M24 SPS models with and without added nebular emission that were presented in this work can be found at `https://sophie-newman.github.io/cloudy-maraston.html`. Here you can also find a set of JUPYTER[7] notebooks that have been created to allow others to recreate the plots in this paper with ease.

(ii) The open source code for SYNTHESIZER can be found in this repository and the GRID-GENERATION open source code can be found here.

Any additional information is available upon reasonable request to the corresponding author.

[7] `https://jupyter.org/`

## APPENDIX A: MASS-LOSS

In this section, we briefly discuss the impact of adjusting the mass-loss parameter in the M13 models. We compare two models, the standard '1 $\dot{M}_{\odot}$' model for solar metallicity and the '2 $\dot{M}_{\odot}$' model where everything is identical but there is twice the amount of mass-loss in massive stars.

The amount of ionizing photons produced by these models are shown in Fig. A1. We see that the two mass-loss models produce a very similar number of photons per second that are capable of ionizing hydrogen with the higher mass-loss model producing slightly more for ages less than around 5.8 Myr. Then the lower mass-loss rate model dominates briefly around 5.8 Myr before the two models align even more closely.

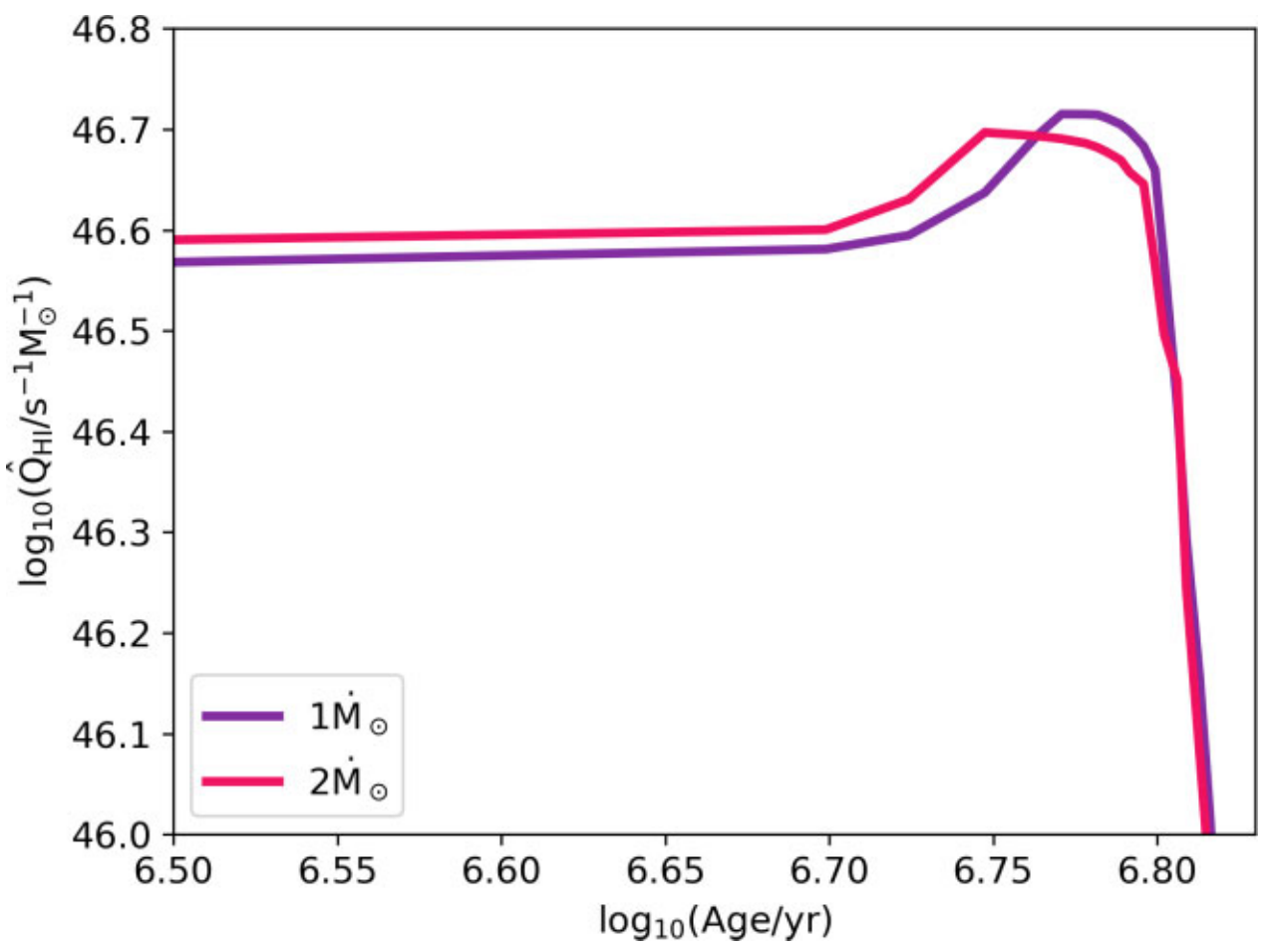

**Figure A1.** The specific ionizing photon luminosity of the M13 model as a function of age for solar metallicity for two different mass-loss models: the standard '1 $\dot{M}_{\odot}$' model for solar metallicity and the '2 $\dot{M}_{\odot}$' model where everything is identical but has twice the amount of mass-loss in massive stars.

The differences between the two mass-loss rate models are small compared to those seen in Fig. 2 (and Fig. 14), so we do not analyse differences in emission line properties due to this effect in this work.

## APPENDIX B: ABUNDANCES

In this work, we adopt the Galactic Concordance abundance framework from D. C. Nicholls et al. (2017), which provides a reference pattern of element abundances based on extensive stellar data in the Milky Way. Table B1 lists the reference logarithmic abundances, defined as $\log_{10}(N_{\rm element}/N_{\rm H})$, for a reference metallicity of $Z = 0.015$. These reference values are consistent with solar composition for the lightest elements and serve as the baseline for scaling other elements with metallicity.

For elements that vary with metallicity, abundances are scaled relative to oxygen using the prescriptions described by D. C. Nicholls et al. (2017). Table B2 presents the logarithmic abundances of these elements for the range of metallicities used in this work. Carbon and nitrogen are treated with special scaling formulas to account for their distinct nucleosynthetic origins, while other elements follow a simpler linear scaling with plateaus at low and high oxygen-to-hydrogen ratios. The resulting carbon and nitrogen abundances for the same metallicity grid are listed within Table B2.

In addition to metallicity-dependent abundances, we consider the depletion of elements onto dust grains. Depletion reduces the gas-phase abundance of an element relative to its total abundance. We adopt the E. B. Jenkins 2009 depletion pattern as implemented in CLOUDY by C. M. Gunasekera et al. (2022), which parametrizes the depletion factor $D_x$ for each element as a function of a scaling parameter $F_*$. In this work, we assume $F_* = 0.5$ to represent a moderate level of depletion. Table B3 lists the resulting depletion factors for the elements considered. Elements with higher $D_x$ are less depleted, whereas lower values indicate stronger incorporation into dust. These depletion factors are applied to compute the gas-phase abundances used in our photoionization models.

**Table B1.** Reference ($Z = 0.015$) logarithmic abundances, $\log_{10}(N_{\rm element}/N_{\rm H})$ from D. C. Nicholls et al. 2017.

| Element | Reference abundance |
|---|---|
| H | 0.00 |
| He | −1.09 |
| Li | −8.722 |
| Be | −10.68 |
| B | −9.193 |
| C | −3.577 |
| N | −4.21 |
| O | −3.24 |
| F | −7.56 |
| Ne | −3.91 |
| Na | −5.79 |
| Mg | −4.44 |
| Al | −5.57 |
| Si | −4.50 |
| P | −6.59 |
| S | −4.88 |
| Cl | −6.75 |
| Ar | −5.60 |
| K | −6.96 |
| Ca | −5.68 |
| Sc | −8.84 |
| Ti | −7.07 |
| V | −8.11 |
| Cr | −6.38 |
| Mn | −6.58 |
| Fe | −4.48 |
| Co | −7.07 |
| Ni | −5.80 |
| Cu | −7.82 |
| Zn | −7.44 |

**Table B2.** Logarithmic abundances, $\log_{10}(N_{\rm element}/N_{\rm H})$, for elements which have scaling with metallicity $Z$ from D. C. Nicholls et al. 2017, ordered by atomic number.

| Element | $Z = 0.0003$ | $Z = 0.002$ | $Z = 0.006$ | $Z = 0.014$ | $Z = 0.02$ |
|---|---|---|---|---|---|
| C | −5.72 | −4.82 | −4.19 | −3.63 | −3.37 |
| N | −6.63 | −5.63 | −4.90 | −4.26 | −3.98 |
| F | −9.26 | −8.44 | −7.96 | −7.59 | −7.44 |
| Ne | −5.61 | −4.79 | −4.31 | −3.94 | −3.79 |
| Na | −7.79 | −6.97 | −6.49 | −5.86 | −5.52 |
| Mg | −6.24 | −5.42 | −4.94 | −4.48 | −4.27 |
| Al | −7.37 | −6.55 | −6.07 | −5.61 | −5.40 |
| Si | −6.30 | −5.48 | −5.00 | −4.54 | −4.33 |
| P | −8.79 | −7.97 | −7.49 | −6.68 | −6.22 |
| S | −6.69 | −5.86 | −5.38 | −4.92 | −4.71 |
| Cl | −8.45 | −7.63 | −7.15 | −6.78 | −6.63 |
| Ar | −7.30 | −6.48 | −6.00 | −5.63 | −5.48 |
| K | −8.76 | −7.94 | −7.46 | −7.00 | −6.79 |
| Ca | −7.53 | −6.71 | −6.23 | −5.73 | −5.48 |
| Sc | −10.79 | −9.97 | −9.49 | −8.90 | −8.59 |
| Ti | −8.92 | −8.10 | −7.62 | −7.12 | −6.87 |
| V | −10.31 | −9.49 | −9.01 | −8.20 | −7.74 |
| Cr | −8.58 | −7.76 | −7.28 | −6.47 | −6.01 |
| Mn | −8.78 | −7.96 | −7.48 | −6.67 | −6.21 |
| Fe | −6.68 | −5.86 | −5.38 | −4.57 | −4.11 |
| Co | −9.27 | −8.45 | −7.97 | −7.16 | −6.70 |
| Ni | −8.00 | −7.18 | −6.70 | −5.89 | −5.43 |
| Cu | −10.02 | −9.20 | −8.72 | −7.91 | −7.45 |
| Zn | −9.44 | −8.62 | −8.14 | −7.51 | −7.17 |

**Table B3.** Depletion factors $D_x$ for selected elements at $F_* = 0.5$, based on the E. B. Jenkins 2009 pattern as modified by C. M. Gunasekera et al. 2022.

| Element | Depletion factor $D_x$ |
|---|---|
| Li | $1.535 \times 10^{-1}$ |
| B | $1.000 \times 10^{0}$ |
| C | $6.880 \times 10^{-1}$ |
| N | $7.762 \times 10^{-1}$ |
| O | $7.464 \times 10^{-1}$ |
| Na | $9.473 \times 10^{-3}$ |
| Mg | $1.698 \times 10^{-1}$ |
| Al | $3.266 \times 10^{-2}$ |
| Si | $1.635 \times 10^{-1}$ |
| P | $6.615 \times 10^{-1}$ |
| S | $5.311 \times 10^{-1}$ |
| Cl | $6.705 \times 10^{-1}$ |
| Ar | $4.064 \times 10^{-1}$ |
| K | $1.187 \times 10^{-1}$ |
| Ca | $2.094 \times 10^{-3}$ |
| Ti | $7.880 \times 10^{-3}$ |
| Cr | $2.796 \times 10^{-2}$ |
| Mn | $4.647 \times 10^{-2}$ |
| Fe | $2.586 \times 10^{-2}$ |
| Ni | $2.084 \times 10^{-2}$ |
| Cu | $1.120 \times 10^{-1}$ |
| Zn | $5.710 \times 10^{-1}$ |

## APPENDIX C: BPT DIAGRAM FOR M13

In Fig. C1, we show the BPT diagram for the M13 model which, unlike M24, does not incorporate stellar tracks that include rotation. Similar trends in metallicity are observed as compared to the M24 model in Fig. 12, but we observe that the line strengths of [O III]$\lambda$5007/H$\beta$4861 drop off much quicker as a function of age.

We again see that higher metallicity models of $Z = 0.01 - 0.02$ best match the JADES data (see as the higher contours) at 1 Myr, but then at 3 Myr the $Z = 0.01$ model fits better than the higher metallicity $Z = 0.02$ model.

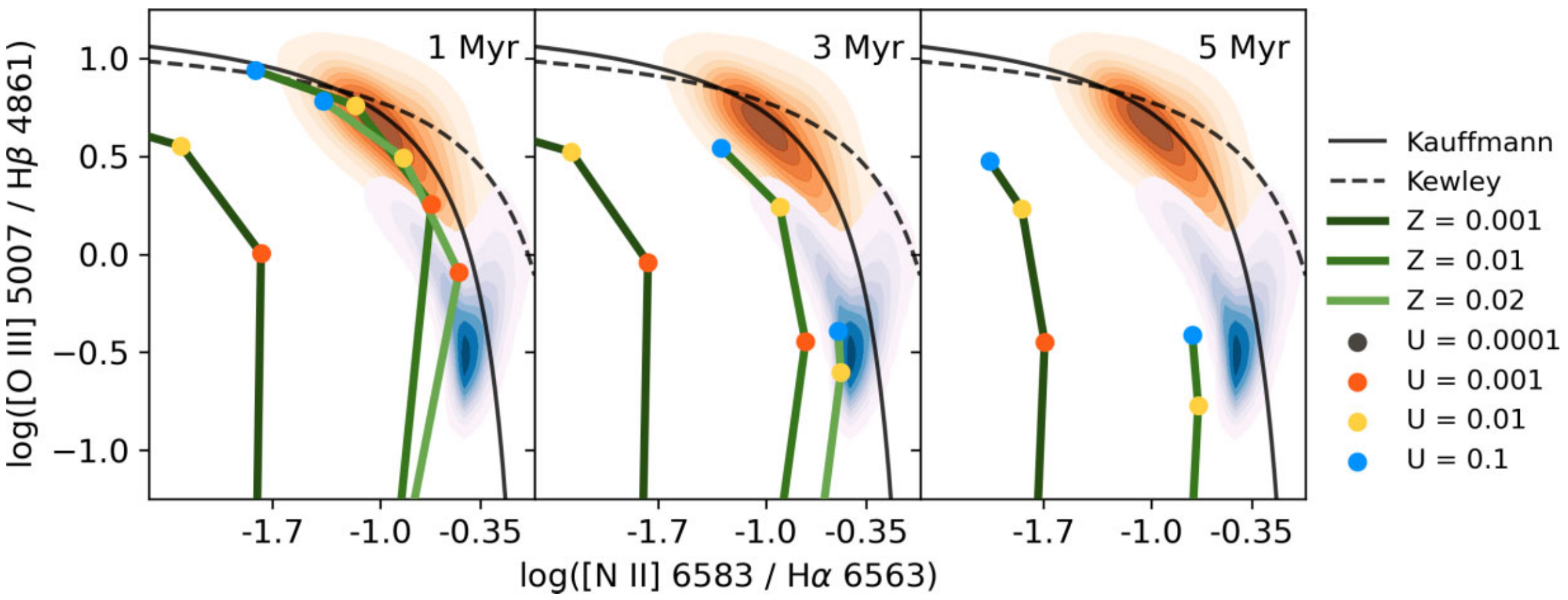

**Figure C1.** Our results for ages 1, 3, and 5 Myr on the Baldwin–Phillips–Terlevich (BPT) diagram, a plot of the ratio [O III]λ5007/H Iλ4863 against the ratio N IIλ6583/H Iλ6561 for the M13 model which, unlike M24, does not include rotation. Each panel corresponds to a different age. Coloured lines represent fixed metallicities, while coloured points indicate varying ionization parameters. Additionally, the Kauffmann and Kewley lines are shown, along with data from SDSS DR8 (shown with the lower contours) and JADES DR3 (shown with the higher contours).

This paper has been typeset from a TEX/LATEX file prepared by the author.